\documentclass[%
 reprint,
superscriptaddress,
 amsmath,amssymb,
 aps,
]{revtex4-1}
\usepackage[utf8]{inputenc}
\usepackage{graphicx}
\usepackage{dcolumn}
\usepackage{bm}
\usepackage{float}
\usepackage{epstopdf}
\usepackage[normalem]{ulem}

\usepackage{braket}
\usepackage{bbm}
\usepackage{xcolor,verbatim}

\begin{document}

\preprint{APS/123-QED}
\inputencoding{latin1}

\title{Feasibility of satellite-to-ground continuous-variable quantum key distribution}

\author{Daniele Dequal}
\email{daniele.dequal@asi.it}
\affiliation{Matera Laser Ranging Observatory, Agenzia Spaziale Italiana, Matera, Italy}
\author{Luis Trigo Vidarte}
\affiliation{Sorbonne Universit\'{e}, CNRS, LIP6, F-75005 Paris, France}
\affiliation{Laboratoire Charles Fabry, IOGS, CNRS, Universit\'{e} Paris Saclay, F-91127 Palaiseau, France}
\author{Victor Roman Rodriguez}
\affiliation{Sorbonne Universit\'{e}, CNRS, LIP6, F-75005 Paris, France}
\affiliation{Thales Alenia Space, F-31100 Toulouse, France}
\author{Giuseppe Vallone}
\affiliation{Dipartimento di Ingegneria dell'Informazione, Universita degli Studi di Padova, via Gradenigo 6B, 35131 Padova, Italy}
\affiliation{Istituto Nazionale di Fisica Nucleare (INFN) - sezione di Padova, Italy}
\author{Paolo Villoresi}
\affiliation{Dipartimento di Ingegneria dell'Informazione, Universita degli Studi di Padova, via Gradenigo 6B, 35131 Padova, Italy}
\affiliation{Istituto Nazionale di Fisica Nucleare (INFN) - sezione di Padova, Italy}
\author{Anthony Leverrier}
\affiliation{Inria Paris, 2 rue Simone Iff, CS 42112, 75589 Paris Cedex 12, France}
\author{Eleni Diamanti}
\email{eleni.diamanti@lip6.fr}
\affiliation{Sorbonne Universit\'{e}, CNRS, LIP6, F-75005 Paris, France}

\date{\today}

\begin{abstract}
Establishing secure communication links at a global scale is a major potential application of quantum information science but also extremely challenging for the underlying technology. While milestone experiments using satellite-to-ground links and exploiting singe-photon encoding for implementing quantum key distribution have shown recently that this goal is achievable, it is still necessary to further investigate practical solutions compatible with classical optical communication systems. Here we examine the feasibility of establishing secret keys in a satellite-to-ground downlink configuration using continuous-variable encoding, which can be implemented using standard telecommunication components certified for space environment and able to operate at high symbol rates. Considering a realistic channel model and state-of-the-art technology, and exploiting an orbit subdivision technique for mitigating fluctuations in the transmission efficiency, we find positive secret key rates for a low-Earth-orbit scenario, while finite-size effects can be a limiting factor for higher orbits. Our analysis determines regions of values for important experimental parameters where secret key exchange is possible and can be used as a guideline for experimental efforts in this direction.
\end{abstract}

\maketitle


\noindent {\large \textbf{Introduction}}\\
Quantum key distribution (QKD) exploits fundamental principles of physics to exchange cryptographic keys between two parties. It can guarantee information-theoretic security,
in the sense that the security of the protocol does not depend on the complexity of some mathematical problem and hence the computational power of a possible adversary does not have to be bounded. QKD represents today one of the most successful applications of quantum information~\cite{Scarani2009, Diamanti16}.

The rapid evolution in QKD implementations has resulted in extending the communication range from few centimeters of the first test to several hundreds of kilometers obtained with modern technology~\cite{Jouguet2013a, Tang2014, Yin2016, Boaron2018}. However, this evolution in ground-based implementations faces a fundamental limitation related to the attenuation of the quantum signal in optical fibers, which increases exponentially with the distance. With this scaling law, covering several thousands of kilometers, as required for the realization of an intercontinental QKD link, would be impossible even with the most advanced technology, if only repeaterless architectures are considered \cite{Pirandola2017}. Quantum repeaters \cite{Briegel1998,Jiang2009, Sangouard2011,Azuma2015, Vinay2017}, whose functioning relies on entanglement distribution and in most cases on quantum memories, might solve the problem of extending the communication range. However, despite progress in the field \cite{Li2019, Bhaskar2020}, the technology is still far from being applicable to intercontinental quantum communication.

To overcome this limitation, a possible solution is the use of orbiting terminals to distribute cryptographic keys among ground stations. Studies investigating the feasibility of quantum communication using satellites have been ongoing for a decade~\cite{Villoresi2008, Yin2013a, Vallone2015, Dequal2016, Carrasco-Casado2016, Takenaka2017, Agnesi2018}, but a milestone was reached recently with the first complete satellite-to-ground QKD implementations realized with the Chinese satellite \textit{Micius}~\cite{Liao2017, Yin2017a}. Soon after these demonstrations, the satellite was used for the realization of the first intercontinental quantum-secured communication~\cite{Liao2018}, thus opening the era of satellite QKD.

While these results represent a major step in the field, several issues still need to be addressed for the realization of a global QKD network based on satellite communication. In this framework, an important aspect is related to the development of high performance space-qualified terminals that will allow for stable, high throughput QKD links from a constellation of satellites to a network of ground stations. To this end, a possible breakthrough may come from the implementation of continuous-variable QKD protocols (CV-QKD)~\cite{Ralph2000, Grosshans2003, Diamanti2015b, Laudenbach2017a}. These protocols have the main advantage of using standard telecommunication components, such as IQ or amplitude and phase modulators for state preparation and coherent receivers for state detection, thus allowing to exploit the heritage of classical optical communication both in terms of high speed components and of their space qualification. The possibility of free-space and satellite CV-QKD has been investigated  theoretically~\cite{Vasylyev2012, Semenov2012, Wang2018, Ruppert2019} and some preliminary experimental studies have been performed on signal transmission along free-space and satellite-to-ground links~\cite{Heim2014, Gunter2017}. 
Moreover, recent studies have summarized the main characteristics of a satellite-based CV-QKD link \cite{Hosseinidehaj2019} and have provided the secret key rate for some specific scenarios, which however do not include crucial link aspects, such as beam divergence, satellite pointing error, satellite motion and finite size effects~\cite{Guo2018, villasenor2020atmospheric, kish2020feasibility}. Therefore, whether this technology can be used for secret key generation in a realistic satellite-based scenario remains an open question.

\begin{figure}
\centering
\includegraphics[width=0.5\textwidth]{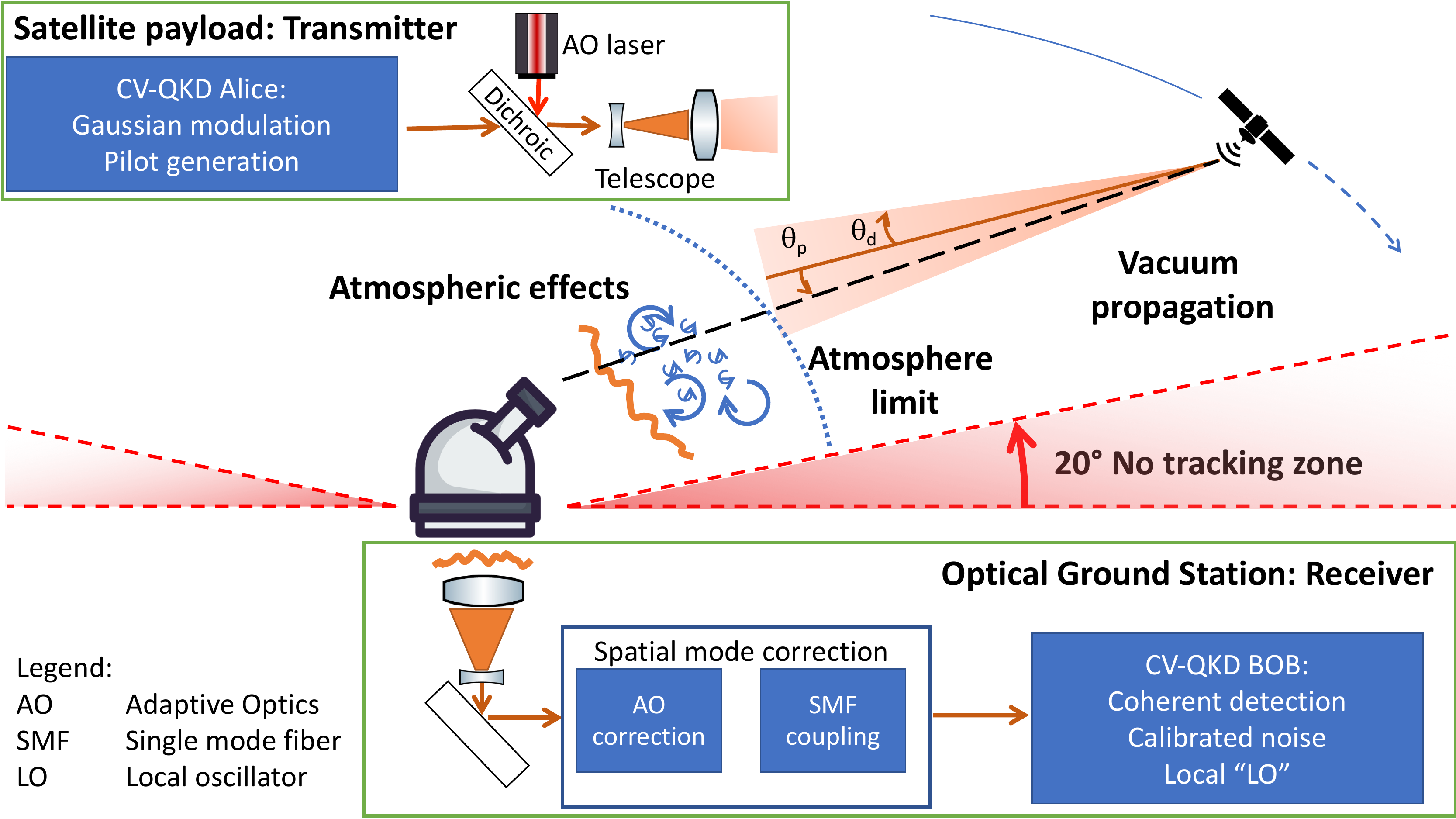}
\caption{Schematic diagram of the CV-QKD communication scheme analyzed in this work. A fixed ground station (Bob) follows the trajectory of a satellite (Alice), equipped with a tracking system, passing over its zenith point. An adaptive optics (AO) system is required in order to correct the wavefront distortions due to the atmosphere and maximize the signal to noise ratio at the receiver. The exact implementation of this system is beyond the scope of this work. The parameters $\theta_p, \theta_d$ are explained in the text.
}
\label{Sketch}
\end{figure}

Here we present a feasibility study of satellite-to-ground CV-QKD, taking into consideration state-of-the-art technology for the quantum state generation, transmission and detection, a realistic channel model and various orbit configurations. Our analysis follows the trusted node approach, where the satellite establishes a separate QKD link with each ground station and hence has access to the keys~\cite{Liao2017, Liao2018}, rather than the untrusted one, where entangled photons are provided by the satellite to the ground stations, which subsequently establish the secret key~\cite{Yin2020}. 
Furthermore, we calculate the secret key rate in the downlink scenario, where the emitter is on the satellite and the receiver on the ground, as it is more favorable for the optical signal transmission. The receiver uses a coherent detector with a free running local oscillator (local LO) and reference symbols (pilots) are transmitted for phase recovery.
A schematic diagram of the configuration we are considering is shown in Fig.~\ref{Sketch}.

Adopting a technique based on orbit subdivision to mitigate the effects of transmission fluctuations, we find that continuous-variable technology is a viable option for satellite QKD on low Earth orbits and identify experimental parameter regions that allow for secret key exchange. For higher orbits, the key generation is affected by finite size effects due to the limited number of symbols exchanged in a single satellite pass for such high-loss channels. These may be mitigated by achieving higher transmission rates or by considering multiple satellite passes.\\


\noindent {\large \textbf{Results}}\\
In our study we first provide a general model of the satellite-to-ground transmission channel, taking into account the beam propagation as well as the satellite orbit. We then examine the effect of channel fluctuations in CV-QKD and derive an equation for the secret key rate over generic fading channels. We subsequently use this equation for estimating the key rate in the case of downlink transmission, both in the asymptotic limit and considering finite size effects.

\medskip

\noindent\textbf{Channel model.} 
We start our analysis by investigating the statistical properties of the satellite-to-ground transmission channel, which are critical for the assessment of the possibility to establish a QKD link in this configuration. In the downlink scenario that we are considering here, the beam travels from the satellite to the ground station and undergoes the disturbance and loss effects due to the atmosphere at the end of its path, resulting in a more favorable situation for key generation with respect to an uplink configuration~\cite{Bourgoin2013}. Indeed, in downlink most of the beam propagation occurs in vacuum, where the beam maintains its diffraction limit properties, while the turbulent atmosphere is encountered only during the last $\simeq20$~km of its path. On the contrary, for uplink the wavefront is distorted at the beginning of its path, resulting in a stronger impact on the beam propagation.

There are several disturbance effects that occur during beam propagation, which can be classified as systematic or of random nature.

The systematic effects are theoretically predictable physical processes that perturb and attenuate the signal, and they include the refraction of the beam in the different atmospheric layers and the extinction of light due to absorption or scattering by air molecules or aerosols. The former is due to the variation in the optical refractive index of the atmosphere as a function of altitude and it causes the light to deviate from a straight line, resulting in an elongation of its physical path. Reference~\cite{Vasylyev2019} provides a detailed calculation of the elongation factor - the ratio of the total length of the beam trajectory to the geometric path length - as a function of the apparent elevation angle of the satellite, \emph{i.e.}, the angle with respect to the horizon at which the satellite appears due to refraction and which differs from the real elevation angle. In this work, we restrict our analysis to elevation angles above 20 degrees, where the elongation factor remains close to 1 and therefore this effect can be neglected (see Fig.~\ref{Sketch}).
The latter effect, namely extinction due to absorption and scattering, depends on the link length and on the molecule and aerosol distribution model~\cite{Vasylyev2019}. It also strongly depends on the sky condition and the transmission wavelength.  For elevation angles above 20 degrees, the atmospheric transmission efficiency $\tau_{\text{atm}}$ scales as:
\begin{equation}
    \tau_{\text{atm}}=\tau_{\text{zen}}^{sec(\theta_{\text{zen}})},
    \label{eq:teff}
\end{equation}
where $\theta_{\text{zen}}$ is the zenith angle and $\tau_{\text{zen}}$ is the transmission efficiency at zenith~\cite{Tomasi2014}.
The estimation of the zenith transmission efficiency relies on the MODTRAN code~\cite{Berk2014}, a widely used atmospheric transmittance and radiance simulator. Considering a 1550~nm wavelength, mid-latitude summer atmospheric model with visibility of 23~km (corresponding to clear sky condition) the MODTRAN web app calculator gives $\tau_{\text{zen}}=0.91$ for both rural and urban aerosol models~\cite{ Modtran}.

In addition to such systematic effects, random variations in the atmospheric temperature lead to fluctuations in the refractive index that have the statistical properties of turbulent scalar fields. The most important consequence of this atmospheric turbulence are intensity fluctuations (scintillation), beam wandering and beam broadening, which induce fading, namely fluctuations in the received optical power and hence in the transmissivity of the channel. The strength of these effects also depends on the altitude and hence on the elevation angle, as discussed in detail in Ref.~\cite{Vasylyev2019}. The atmospheric turbulence is also responsible for the deformation of the beam profile. This is crucial, especially in the context of CV-QKD, where mode matching between the received signal and the phase reference (local oscillator) is important for the coherent detection~\cite{Gunter2017}. To avoid mode mismatch, we assume the use of single mode fibers as spatial-mode filters of the incoming beam, together with an advanced adaptive optics system~\cite{Tyson2011} to improve the coupling efficiency of the incoming light into the single mode fiber core. Acting as a spatial mode filter, the coupling to a single mode fiber removes components of the signal that would not interact with the LO and contribute to the detected signal. This filtering hence reduces the noise in the detection apparatus and also facilitates the use of components like integrated coherent receivers, which are typically available as commercial off-the-shelf and standardized devices.
We remark that recent advances in this field have experimentally demonstrated a coupling efficiency in a single mode fibre exceeding $50\%$ for a large aperture telescope~\cite{Jovanovic2017a}.

Besides turbulence effects, the beam propagation is affected by wandering due to the pointing error of the satellite. This is characterized by the angle $\theta_p$, which is defined as the standard deviation of the angle between the direction of the center of the beam and the imaginary line joining the emitter and receiver telescopes, so that in the case of no pointing error we would have $\theta_p=0\ \mathrm{\mu rad}$. A pointing error of the order of 1 $\mu$rad has been obtained in low-Earth-orbit (LEO) satellite-to-ground communication links \cite{Liao2017}. This is used as a nominal value in our analysis. Similarly, the divergence of the beam is characterized by the angle $\theta_d$, for which we use the nominal value of 10~$\mu$rad, which has  been demonstrated with a 300~mm aperture telescope on-board of the \textit{Micius} satellite.

We are now ready to analyse the statistical properties of our channel, which will be necessary for assessing the effect of fading on the CV-QKD link, under the above assumptions. To do this, we follow the approach of Ref.~\cite{Vasylyev2012} and calculate the probability distribution of the transmission efficiency (PDTE), as it characterizes completely the statistics of the quantum channel for a given satellite orbit. Indeed, the transmission of coherent states of light through the atmosphere can be modeled by the input/output relation of the  annihilation operators, $\hat{a}_{\mathrm{out/in}}$. The transformation should preserve the commutation relation, so that we can write:
\begin{equation}
    \hat{a}_{\mathrm{out}} = T\hat{a}_{\mathrm{in}} + \sqrt{1-T^2}\hat{c},
\end{equation}
where $\hat{c}$ are environmental modes and $T$ is the transmission coefficient (with the transmission efficiency being $\tau=T^2$). Within this model, we can obtain the $P$-function characterizing the statistics of the quantum state; it is then possible to show that the PDTE is sufficient to characterize the state at the receiving telescope \cite{Vasylyev2012}.
In the following, we first calculate the probability distribution obtained at a fixed distance between the satellite and the ground station, and then we take into account the satellite's orbit to compute the total probability distribution, \emph{i.e.}, the PDTE of the entire orbit.

\medskip

\paragraph{Probability distribution at a fixed satellite distance:}
We consider a fixed distance $R$ between the satellite and the ground station. The overall transmission efficiency can be divided into a fixed and a time varying term. We estimate the fixed attenuation term to be 3.8 dB, including 3 dB of losses for fiber coupling and an additional 0.8 dB for taking into account the fact that we are only considering the main peak of the Airy diffraction pattern. As discussed previously, the main dynamic effects affecting the transmission in our analysis are the pointing error of the satellite and the divergence of the beam, characterized by the angles $\theta_p$ and $\theta_d$, respectively.

\begin{figure}
\centering
\includegraphics[width=0.5\textwidth]{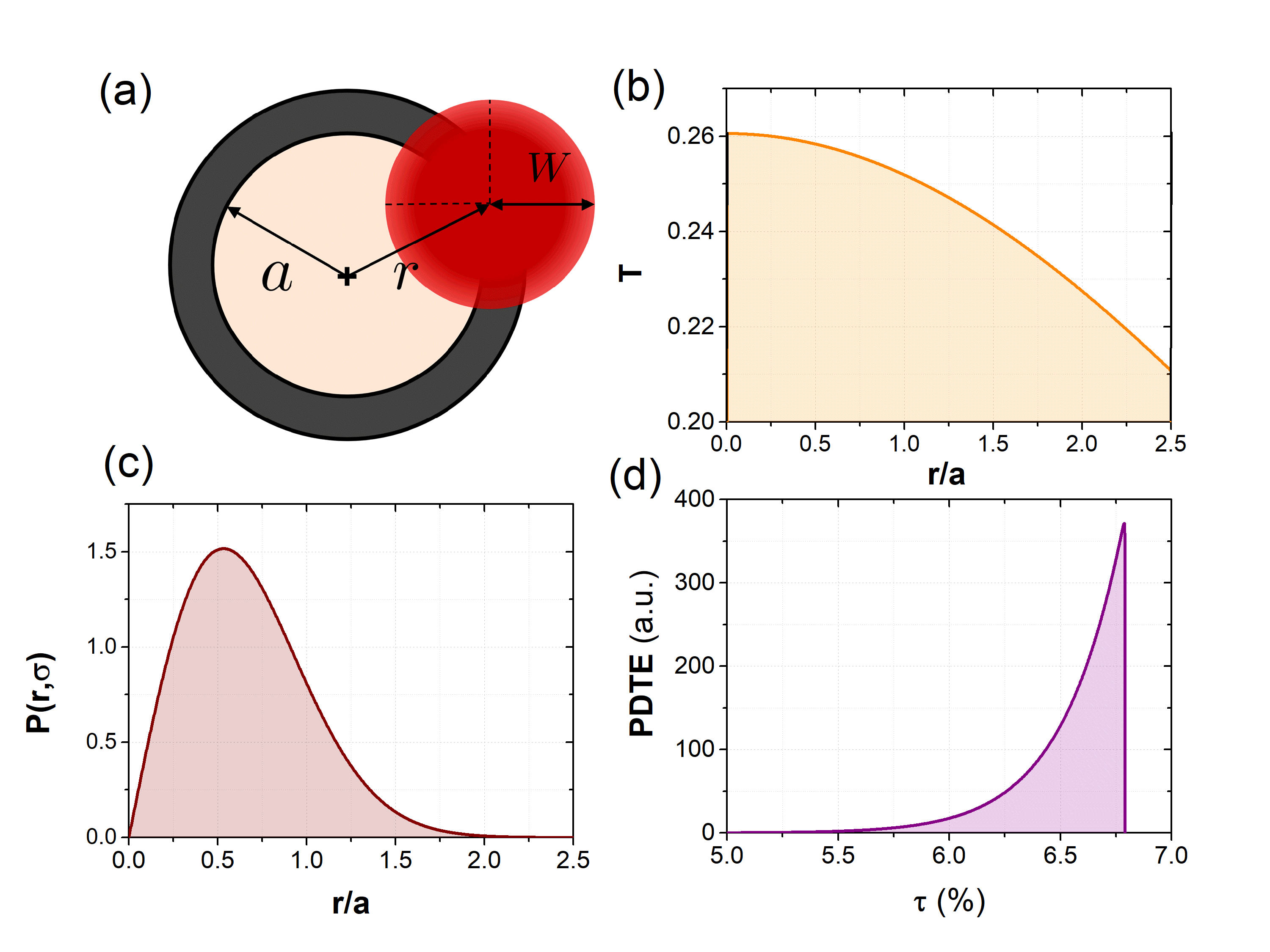}
\caption{Example of the characterization of an atmospheric channel at a fixed satellite-to-ground slant distance of $R=400$ km. The values for the variables are $\theta_p=1\ \mu$rad, $\theta_d=10\ \mu$rad, {$a=0.75$} m. (a) Schematic of the beam and receiver telescope aperture; (b) Transmission coefficient as a function of the deflection distance; (c) Probability distribution of the deflection distance; (d) Probability distribution of the transmission efficiency (PDTE).}
\label{Static_Weib}
\end{figure}

Following Ref.~\cite{Vasylyev2012}, to calculate the PDTE we first consider the deflection distance, $r$, and its standard deviation, $\sigma_r$. As shown in Fig.~\ref{Static_Weib}(a), $r$ is the instantaneous distance between the center of the receiving telescope and the center of the beam. Its standard deviation depends on the pointing and on the atmospheric turbulence as:
\begin{equation}
    \sigma_r = \sqrt{(R\theta_p)^2 + \sigma_{\mathrm{turb}}^2}\simeq R\theta_p.
    \label{sigma_sum}
\end{equation}

In the weak turbulence regime, the variance of the beam center due to turbulence is given by $\sigma_{\mathrm{turb}}^2 \simeq 1.919\; C_n^2 z^3 (2W_0)^{-1/3}$  and depends on the distance traveled by the beam in the atmosphere, $z$, and on the beam waist when entering the atmosphere, $W_0$. For stronger turbulence, this expression represents an upper bound, as $\sigma_{\mathrm{turb}}^2$ saturates and an increase of the path length or turbulence strength will not increase its value \cite{Fante1980}. The parameter $C_n^2$ is the refractive index structure parameter that characterizes the strength of the atmospheric turbulence. In case of moderate turbulence and considering a wavelength of 1550~nm we have  $C_n^2 \simeq 10^{-15} - 10^{-14}\ \mathrm{m}^{-2/3}$, which gives $\sigma_{\mathrm{turb}}^2$ $\lesssim$ $10^{-4}\ \mathrm{m}^2<<(R\theta_p)^2 \simeq 10^{-1}\ \mathrm{m}^2$, corresponding to a pointing error of $\simeq 1\ \mathrm{\mu rad}$ and a satellite altitude of 300 km. This justifies the approximation in the right hand side of Eq.~(\ref{sigma_sum}) for all satellite altitudes above 300~km.

Under this approximation, the probability distribution of the deflection distance follows the Weibull distribution:
    \begin{equation}
    P(r; \sigma_r) =\frac{r}{\sigma_r^2}\exp\left(-\left(\frac{r}{\sqrt{2}\sigma_r}\right)^2\right).
    \label{P-r}
\end{equation}
An example of this distribution is shown in Fig.~\ref{Static_Weib}(c). 
Given now a distance $r$, the transmission coefficient can be obtained from geometrical considerations. An approximate but sufficiently accurate analytic relation between $r$ and $T$ can be calculated as~\cite{Vasylyev2012}:
\begin{equation}
    T^2(r) = T_0^2\exp\left(-\left(\frac{r}{S}\right)^{\lambda}\right).
    \label{T-r}
\end{equation}

$T_0$ is the maximum transmission coefficient possible, and $S$ and $\lambda$ are the scale and shape parameters respectively,  given by:
\begin{align}
    S=a \left[ ln\left(\frac{2 T_0^2}{1-exp[-4 \frac{a^2}{W^2}] I_0(4\frac{a^2}{W^2})}\right)\right]^{-(1/\lambda)},
    \label{Scale_para}
\end{align}

\begin{align}
\begin{split}
    \lambda &=8\frac{a^2}{W^2}\frac{exp[-4\frac{a^2}{W^2}] I_1(4 \frac{a^2}{W^2})}{1-exp[-4\frac{a^2}{W^2}] I_0(4 \frac{a^2}{W^2})}\\
    &\times \left[ ln\left(\frac{2 T_0^2}{1-exp[-4 \frac{a^2}{W^2}] I_0(4\frac{a^2}{W^2})}\right)\right]^{-1}, 
    \label{Shcape_para}
\end{split}
\end{align}
where $I_n$ is the $n$-th order modified Bessel function.

All three are given functions of the beam waist on the ground, $W=R\theta_d > 4$~m for satellites above 400 km, and of the telescope aperture radius, $a$, here considered 0.75~m. Hence, we can write $T_0=T_0(W,a)$, $\lambda = \lambda(W,a)$, and $S = S(W,a)$. The relation between $T$ and $r/a$ for these values is shown in Fig.~\ref{Static_Weib}(b).

We can then substitute Eq.~(\ref{T-r}) into Eq.~(\ref{P-r}) and use the chain rule to obtain the probability distribution of the transmission coefficient, PDTC. The PDTE is obtained from the PDTC using the chain rule with $\tau = T^2$. Fig.~\ref{Static_Weib}(d) gives an example of the characterization of an atmospheric channel of fixed distance following our model for the same parameters as discussed above.

\medskip

\paragraph{Probability distribution for orbit:}
We now obtain the PDTE for the entire satellite pass. In our analysis, we consider circular orbits that are passing at the zenith of the ground station (which is assumed not to move during the pass).  We can write the radius of such orbits as $R_O =R_E + h_s$ , were $R_E$ is the Earth's radius and $h_s$ the satellite's altitude with respect to the ground. The angular velocity of the satellite is $\omega^2 = G M_T / R_O^3$, where $M_T$ is the Earth's mass and $G$ is the gravitational constant. The distance between the satellite and the ground station during the satellite's visibility time, that we denote $R(t)$, then reads:

\begin{equation}
    R(t)=\sqrt{R_E^2+R_O^2-2R_E R_O\cos(\omega t)}.
    \label{traj}
\end{equation}

\begin{figure}
\centering
\includegraphics[width=0.5\textwidth]{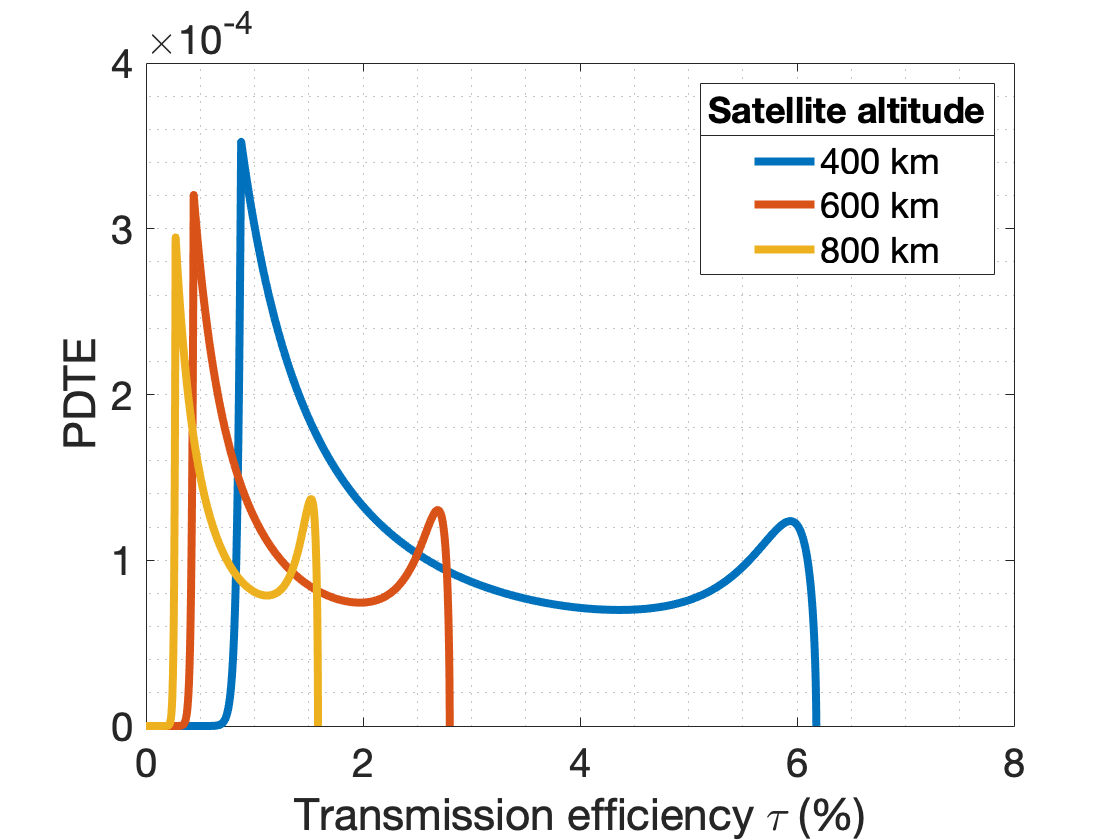}
\caption{PDTE for three different orbits of increasing satellite altitude. The values of the variables for each orbit are the same as in Fig. \ref{Static_Weib}. 
}
\label{orbits}
\end{figure}

We then proceed as follows:

 \begin{itemize}
   \item  The orbit is divided into a set of points defined by the position of the satellite at a certain time, $R(t_i)$ ($i$ runs with the number of points), given by the orbital equation, Eq.~(\ref{traj}).
   \item For each one of these points, both the PDTE$(R(t_i))$ and the time difference between consecutive points of the orbit, denoted  $\Delta t_i=t_i-t_{i-1}$, are computed. The PDTE$(R(t_i))$ includes as a multiplicative factor the atmospheric transmission efficiency, Eq.~(\ref{eq:teff}), for the elevation angle corresponding to $R(t_i)$. The value PDTE$(R(t_i))\cdot\Delta t_i$ gives the distribution of the times with different transmission efficiencies inside the computed interval.
   \item Therefore, if we sum PDTE$\cdot\Delta t_i$ over all the points we obtain the final distribution for the time spent by the satellite with a certain transmission efficiency $\tau$. Indeed, we are mimicking the integral over the flight time:
   \begin{equation}
       \frac{1}{\mathrm{FT}}\sum_{i}\mathrm{PDTE}(\tau,R(t_i))\Delta t_i \longrightarrow \frac{1}{\mathrm{FT}}\int\mathrm{PDTE(\tau,t)\mathrm{d}t},
   \end{equation}
\end{itemize}
where the flight time, FT, is the normalization factor.
Because we are considering circular orbits, we can label each orbit with its altitude, which is the minimum distance of the orbit, coinciding with the moment at which the satellite is exactly above the ground station. For such orbits and following the procedure described above, we show in Fig.~\ref{orbits} the probability distribution of the transmission efficiency (PDTE) for three different orbits of increasing altitude for a telescope with aperture radius $a=0.75$ m.
We remark that for higher orbits the variance of the distribution decreases. As described in the following, this fact has an impact on the noise introduced in time varying channels.

\begin{figure}
\centering
\includegraphics[width=0.5\textwidth]{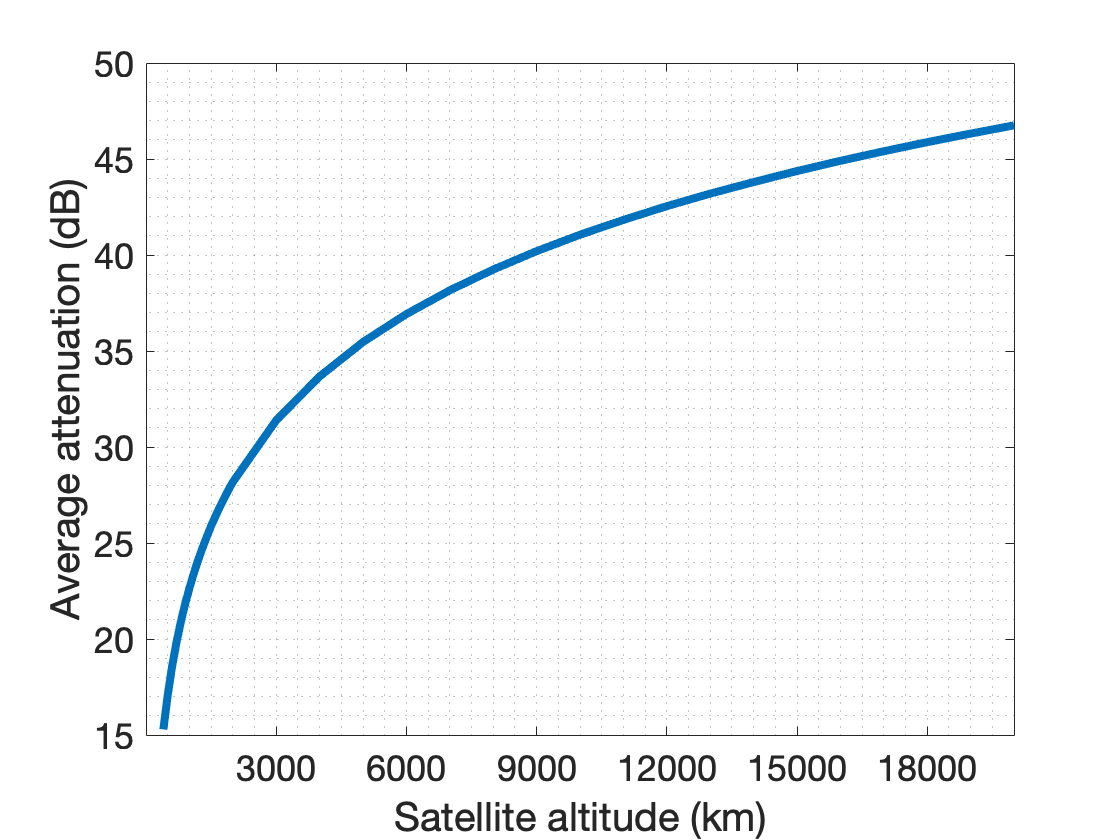}
\caption{Average attenuation per pass as a function of satellite altitude.}
\label{orbit_attenuation}
\end{figure}

We note that the conclusions that we have drawn for the downlink characterization are in agreement with the recent analysis of Ref.~\cite{Liorni2019}. Interestingly however the authors there use the elliptical model rather than the circular one, which means that the ellipticity does not affect the probability distributions. For completeness, we also show in Fig.~\ref{orbit_attenuation} the average attenuation encountered in a pass as a function of the satellite altitude.

\medskip


\noindent\textbf{Key rate estimation.} 
Let us now describe the procedure we follow to estimate the key rate over a fading channel in the asymptotic regime, \emph{i.e.}, when no finite-size effects are taken into account. For this estimation we consider the no-switching CV-QKD protocol~\cite{Weedbrook2004} in its prepare and measure version (PM).
Alice starts by sampling $2N$ real random variables $X_1, ..., X_{2N}$ according to a Gaussian distribution with variance $V_A$, that is, $X_k \sim \mathcal{N}(0, V_A)$ and prepares the corresponding $N$ coherent states $\ket{\alpha_1}, . . . , \ket{\alpha_N}$, where $\alpha_k = X_{2k-1} + i X_{2k} \in \mathbbm{C}$. Each of these states is sent through the quantum channel to Bob, who performs measurements in both quadratures simultaneously (heterodyne detection) \footnote{In practice, Bob splits the signal onto a balanced beamsplitter then measures the $\hat{q} = b+b^\dagger$ quadrature of one output mode and the $\hat{p} = i(b^\dagger - b)$ quadrature of the second output mode. He then stores the first measurement outcome in the variable $Y_{2k-1}$ and the second outcome in $Y_{2k}$.}. For the $k^{\mathrm{th}}$ use of the channel, he obtains two results $Y_{2k-1}$ and $Y_{2k}$ which are supposed to be correlated to $X_{2k-1}$ and $X_{2k}$. The string $\mathbf{Y} =(Y_1, \ldots Y_{2N})$ forms the raw key since we consider the reverse reconciliation setting~\cite{Grosshans2002Arxiv} which is advantageous in case of low transmission efficiency. Note that in a practical protocol, Bob will discretize his data, for instance by dividing the real axis into bins of small width. Asymptotic values are obtained in the limit $N \to \infty$.

The standard formula to compute the asymptotic value of the secret key rate, in the case of reverse reconciliation, is the so-called Devetak-Winter bound~\cite{Devetak2005}:
\begin{align}
\label{DW_rate}
K_{\text{DW}} = \beta I_{AB} - \chi_{BE},
\end{align}
where $\beta I_{AB}$ quantifies the correlations between Alice and Bob's data (here, the imperfect efficiency of the error correction procedure is taken into account thanks to parameter $\beta \leq 1$) and $\chi_{BE}$ quantifies how much information the adversary holds about the raw key corresponding to Bob's string. The Devetak-Winter bound is valid against collective attacks and remains true even against general attacks for QKD protocols with sufficient symmetry, including for the no-switching protocol, more precisely when de Finetti reductions are applicable~\cite{Renner2009, Christandl2009, Leverrier2017}.

In order to assess the performance of a protocol for a given quantum channel, one simply needs to estimate the value of $\beta I_{AB}$ and $\chi_{BE}$. For the first term, since we are dealing with the reverse reconciliation scenario, one should provide a model of the classical channel $\{Y_{k} \to X_{k}\}$ as well as an error correction procedure allowing Alice to recover the value of $Y_k$ from her observations and from additional side-information sent by Bob.
In order to obtain $\chi_{BE}$, one should similarly model the parameter estimation procedure and compute the expected value that Alice and Bob would observe for our specific channel model.
While these computations are fairly standard in the case of a fixed Gaussian channel with constant transmission efficiency and excess noise, the situation becomes more subtle in the case of a fading quantum channel and indeed conflicting results have appeared in the literature~\cite{Usenko2012, Papanastasiou2018} (see Methods for details).

Here, we find it useful to recall the derivation of the asymptotic secret key rate from the non-asymptotic case. According to Refs.~\cite{Leverrier2015a, Leverrier2017}, the protocol we are considering is secure against general attacks, even in the finite size regime, and the asymptotic secret key rate is given~by
\begin{equation}
\label{eq_key_general}
    K=\lim_{N \to \infty} \frac{1}{N}\left( H(\mathbf{Y} ^{(N)}) - \mathrm{leak}_{\mathrm{EC}}^{(N)}) \right) - f(\Gamma^{(N)}).
\end{equation}
In this expression, $ H(\mathbf{Y}^{(N)})$ refers to the empirical entropy of the string $\mathbf{Y} ^{(N)}$ and the superscript $N$ is explicitly written to emphasize that each of these quantities depends on the block length. Since we are only interested in the asymptotic behaviour of the secret key rate, we neglect discretization effects here. The quantity $\mathrm{leak}_{\mathrm{EC}}^{(N)} $ is the number of bits that are leaked in the error correction procedure during which Bob sends some side information to Alice to help her guess the value of $\mathbf{Y}$. The term $f(\Gamma^{(N)})$ quantifies the information available to Eve and will be described later.

The advantage of Eq.~\eqref{eq_key_general} is that it tells us how to compute $\beta I_{AB}$ and $\chi_{BE}$ in the Devetak-Winter bound, namely
\begin{equation}
        \begin{aligned}[b]
        \beta I_{AB} &= \lim_{N \to \infty} \frac{1}{N}\left( H(\mathbf{Y} ^{(N)}) - \mathrm{leak}_{\mathrm{EC}}^{(N)}) \right),\\
        \chi_{BE} &= \lim_{N \to \infty}f(\Gamma^{(N)}).
        \end{aligned}
\label{eqn2.qo}
\end{equation}

Let us first consider the first term. Here we model the quantum channel between Alice and Bob as a phase-insensitive noisy bosonic channel with transmission efficiency given by a random variable $\tau_k \in [0, 1]$, whose probability distribution is the one calculated previously. The channel noise will be treated with the so-called excess noise, $\xi$, whose full derivation will be given in the following. We will additionally model the imperfections in Bob's detectors by two parameters: their detection efficiency $\eta$ and the electronic noise $\nu_{\text{el}}$.
In particular, this implies that the random variables $X_k$ corresponding to Alice's inputs and $Y_k$ for Bob's measurement results satisfy:
\begin{align}
Y_k = T_k X_k + Z_k,
\end{align}
where $T_k$ is the overall transmission coefficient for the $k^{\mathrm{th}}$ channel use, $T_k^2=\tau_k$, and $Z_k \sim \mathcal{N}(0,\sigma^2)$ is a Gaussian noise of variance $\sigma^2$ assumed to be constant.

In order to compute the key rate of Eq.~\eqref{eq_key_general}, it is important to understand how fast the fading process is. The main idea here is that this process, whose time scale is typically of the order of 1-10 ms due to atmospheric turbulence, is much faster than the time needed to distill a secret key,  which in our case corresponds to a complete satellite pass. In other words the channel transmission coefficient fluctuates significantly over $N$ uses of the channel, but this coefficient is relatively stable over consecutive uses of the channel, which occur with ns separation. As a consequence, Alice and Bob can exploit classical signals to roughly monitor the current transmission value of the channel and adapt their error correction procedure accordingly. This implies notably that for the error correction procedure, we can assume that Alice and Bob know (approximately) the value of $T_k$. This allows them to use good error correcting techniques developed for the fading channel where the fading process $T_k$ is known to the receiver. In particular, the Gaussian modulation permits to achieve the capacity of this channel up to a reconciliation efficiency factor $\beta$ and one expects~\cite{LapidothShamai2002}
\begin{align}
\label{eq_IAB}
\beta I_{AB} = \beta \mathbf{E}\left[ \log_2 \left( 1+ \frac{T^2 V_A}{\sigma^2} \right) \right],
\end{align}
where $\mathbf{E}[\cdot]$ is the expectation with respect to the fading process. Here and in the following, we write $T$ instead of $T_k$ and replace averages of the form $\frac{1}{N} \sum_{k=1}^N$ by the expectation $\mathbf{E}$ for simplicity.
Note that since the $\log$ function is concave, the value we find for $\beta I_{AB}$ is smaller than the one computed for a channel with a fixed  transmittance $\mathbf{E}[ T^2]$. To numerically compute the value of Eq.~\eqref{eq_IAB} it is possible to use the expressions given in Ref.~\cite{Fossier2009} for a fixed transmission channel, and take their expectation value.

Let us now turn to the second term of Eq.~\eqref{eq_key_general}, namely $f(\Gamma^{(N)})$, which quantifies the information available to Eve. More precisely, $\Gamma^{(N)}$ is a worst case estimate of the (average) covariance matrix of the state Alice and Bob would share in the entanglement-based version of the protocol and
the function $f$ is defined as
\begin{align}
\label{eq_f_Gamma}
    f(\Gamma)= g(\nu_1) + g(\nu_2) - g(\nu_3) - g(\nu_4),
\end{align}
where $g$ is the entropy function $g(z)=\frac{z+1}{2} \log_2\frac{z+1}{2} - \frac{z-1}{2} \log_2\frac{z-1}{2}$, $\nu_1$ and $\nu_2$ are the symplectic eigenvalues of $\Gamma^{(N)}$ and $\nu_3$ and $\nu_4$ are the symplectic eigenvalues of the matrix describing Eve's system conditional on Bob's measurement outcome~\cite{Lodewyck2007}. The interpretation of the function $f$ is that it coincides with the Holevo information between the raw key and Eve's quantum memory computed for a Gaussian state $\rho^G_{\mathrm{ABE}}$ with covariance matrix coinciding with $\Gamma^{(N)}$ on Alice and Bob's systems.

In order to compute the covariance matrix $\Gamma^{(N)}$ that Alice and Bob would infer from their data, we note first that for a fixed transmittance value $T$, the covariance matrix of the bipartite quantum state they would hold in the entanglement-based version of the protocol reads
\begin{align}
\Gamma(T) &= \begin{bmatrix}
    V \mathbbm{1}_2 & T \sqrt{V^2-1}\sigma_Z \\
    T \sqrt{V^2-1}\sigma_Z & (T^2 (V-1) + \sigma^2) \mathbbm{1}_2
    \label{Gamma_Fixed_Channel}
\end{bmatrix},
\end{align}
with $V = V_A+1$, $\mathbbm{1}_2 = \text{diag}(1,1)$ and $\sigma_Z = \text{diag}(1,-1)$.

As observed in Ref.~\cite{Usenko2012}, when the fluctuation of the transmission efficiency is considered, the resulting state is a mixture of the individual fixed-transmission states, giving an overall covariance matrix equal to $\Gamma^{(N)}  = \mathbf{E}[\Gamma(T)]$, that is:
\begin{align}
\label{eq_fading_covariance}
\Gamma^{(N)} &= \begin{bmatrix}
    V \mathbbm{1}_2 & \mathbf{E}[T] \sqrt{V^2-1}\sigma_Z \\
    \mathbf{E}[T] \sqrt{V^2-1}\sigma_Z & (\mathbf{E}[T^2] (V-1) + \sigma^2) \mathbbm{1}_2
\end{bmatrix}.
\end{align}

If we compare the covariance terms in  Eqs.~\eqref{Gamma_Fixed_Channel} and \eqref{eq_fading_covariance} we can identify an effective transmission for the fading channel equal to $\mathbf{E}[T]^2$.
In particular, the variance of Bob's system can be written
\begin{align*}
\mathbf{E}[T^2] (V-1) + \sigma^2 = \mathbf{E}[T]^2 (V-1+ \xi_{\text{fad}}) + \sigma^2,
\end{align*}
where
\begin{align}
\xi_{\text{fad}} = \frac{(\mathbf{E}[T^2] - \mathbf{E}[T]^2)}{\mathbf{E}[T]^2} (V-1) =  \frac{\mathrm{Var}(T)}{\mathbf{E}[T]^2} (V-1)
\end{align}
corresponds to noise exclusively due to fading.
In other words, Eve's information in the presence of fading corresponds to her information for a fixed Gaussian channel with transmission efficiency $\mathbf{E}[T]^2$ and an added noise given by $(V-1) \mathrm{Var}(T)/\mathbf{E}[T]^2$.
This extra noise will be detrimental to the performance of the QKD system unless $\mathrm{Var}(T) \ll \frac{1}{V-1}$. By re-writing the fading case as a fixed case with an effective transmission efficiency and excess noise, it is possible to use the equations reported in Ref.~\cite{Fossier2009} for calculating the eigenvalues in Eq.~\eqref{eq_f_Gamma}.

To summarize, by putting together the two terms of Eq.~\eqref{eq_key_general}, our expression for the secret key rate in the presence of fading becomes:
\begin{align}
\label{key_rate_fading}
K_{\text{fad}} = \beta \mathbf{E}\left[ \log_2 \left( 1+ \frac{T^2 V_A}{\sigma^2} \right) \right] - f (\mathbf{E}[\Gamma (T)]).
\end{align}

\medskip


\noindent\textbf{Simulation results.} 
We are now ready to use the results derived above to estimate the expected key rate achievable for a satellite-to-ground CV-QKD link under our assumptions. To properly account for the expected noise, we include in our modeling the noise contribution related to the phase recovery between the signals generated by Alice and measured by Bob. The technique that we consider here has been proposed in Refs.~\cite{Soh2015, Qi2015} and consists in sending periodic reference symbols (pilots) along with the quantum signal. At the receiver side, Bob uses a free running local oscillator, which must be tuned to compensate for the Doppler frequency shift introduced by the satellite motion, to measure both the pilot and the quantum signals, in a so-called `local' local oscillator configuration. As described in the Methods section, two noise contributions arise from this technique, which are due to laser instability and shot noise. 

We remark that at telecom wavelength, the Doppler shift ranges from several GHz for LEO to several hundreds of MHz for MEO~\cite{DopplerLEO}. This problem is well known in classical laser communication and several solutions have been proposed, such as optical~\cite{Shoji2012} or digital~\cite{Paillier2019} phase-locked loops. An alternative solution could come from precise orbit determination (POD) based on additional satellite payloads, such as retroreflectors, GPS receivers or DORIS antennas. With these techniques it is possible to achieve an \textit{a posteriori} determination of the satellite velocity with a precision of $<1$~mm/s, which would correspond to a residual frequency shift of $<1$~kHz \cite{Gao2015, Kuchynka2020}.  Moreover, in the case of `local' LO CV-QKD, an alternative solution is to exploit the pilots  to measure the residual Doppler shift. In fact, by using ephemeris data, it is possible to pre-compensate the Doppler shift with an \textit{a priori} residual error of tens of MHz, much smaller than the pilot repetition rate. Under these conditions, it would be possible to retrieve the residual Doppler shift by analyzing the trend of the pilot phase. This possible Doppler correction technique will need further experimental investigation, which is however outside the scope of this work.

The overall excess noise $\xi$, here referred to the channel input, is given by the above mentioned contributions, the fading noise, described in the previous section, and an additional fixed contribution due to experimental imperfections, $\xi_{\mathrm{fix}}$, which includes also other possible errors in the phase correction.

\begin{figure}
\centering
\includegraphics[width=0.5\textwidth]{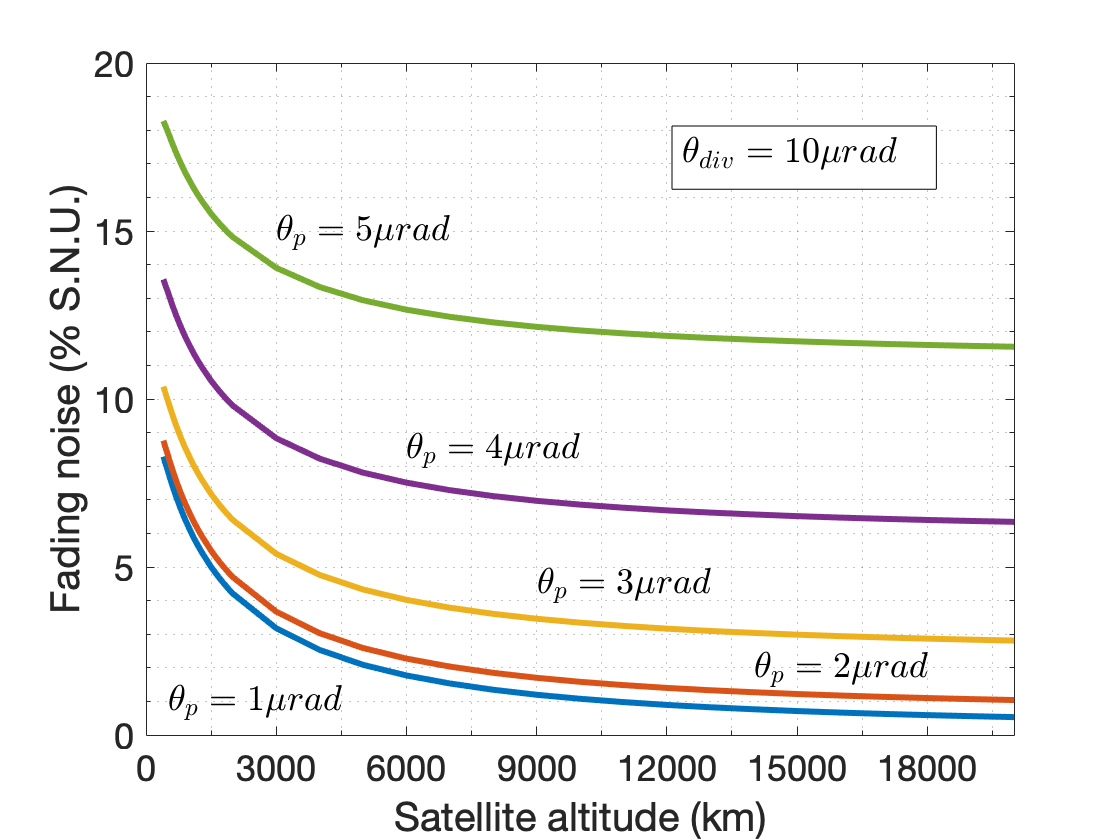}
\caption{Trend of the fading excess noise, $\xi_{\mathrm{fad}}$, in percentage of the shot noise units, as a function of the satellite altitude for several values of pointing error and a fixed value of the divergence angle.}
\label{Scan_fading_noise}
\end{figure}

\begin{table}
    \centering
    \begin{tabular}{|l|c|c|}
    \hline
        \textbf{Parameter} & \textbf{Symbol} & \textbf{Reference value} \\ \hline
        Pointing error & $\theta_p$ & 1 $\mu$rad \\
        Divergence angle & $\theta_d$ & 10 $\mu$rad \\
        Fixed attenuation & Att & 3.8 dB \\
         Zenith transmittance & $\tau_{\text{zen}}$ & 0.91 \\
        Electronic noise & $\nu_{\mathrm{el}}$ & 10\% S.N.U.\\
        Detection efficiency & $\eta $ & $0.4 $\\
        Fixed excess noise & $\xi_{\mathrm{fix}}$ & 1-5\% S.N.U.\\
        Reconciliation efficiency & $\beta$ & 0.95 \footnote{Note that while values of $\beta \geq 0.95$ have been achieved for a Gaussian channel with fixed transmission efficiency \cite{Jouguet2013a} (corresponding to the so-called additive white Gaussian noise channel), some research will be needed to obtain similar performances for fading channels.}\\
        Transmission symbol rate & $f_{\mathrm{TX}}$ & 1 Gsymbol/s\\
        Receiving telescope radius & $a$ & 0.75 m\\
        \hline
    \end{tabular}
    \caption{Summary of the main simulation parameters used in our model, together with their reference values. }
    \label{tab:ref_values}
\end{table}
The main experimental parameters that influence the key rate generation are summarized in Table~\ref{tab:ref_values}, together with their reference values. The reference values considered for the ground station and the satellite are similar to those reported in Ref.~\cite{Liao2017} and represent a high performance satellite optical communication system.  A detailed analysis of the effect of individual parameters on the key rate is given in the Methods. Regarding the signal variance $V_A$, for each satellite altitude and for each set of parameters we choose the value that maximizes the key rate. These values are in general between 2 and 4 shot noise units (S.N.U.), depending on the configuration. Figure~\ref{Scan_fading_noise} shows the fading noise given by the PDTE that we obtain for orbits going from 400 km to 22000 km. As we see, an increase of the noise is present for LEO. This is due to the fact that in such orbits the variation of the slant range is more pronounced thus introducing a higher variance on $\tau$ (as we observe in Fig.~\ref{orbits}). Moreover, it is worth noting that when the pointing error is much smaller than the beam divergence, the fading effect is mainly due to the variation on the satellite distance.

To reduce the effect of fading excess noise, a natural strategy is to reduce the variance of the fading process. This can be achieved as follows: Alice and Bob can approximately monitor the value of the transmission efficiency of the channel seen by the quantum symbols $\tau_k$ by multiplexing in some degree of freedom an intense optical signal that serves as beacon and experiences a transmission efficiency $\tau_b$. An intensity detection of the beacon at Bob's, sampled at rates higher than the atmospheric coherence time (typically $\sim 1\ \mathrm{kHz}$), can provide an accurate estimation of the channel transmittance evolution with time $\tau_b(t)$. This information can be used to classify the detected quantum symbols in groups as a function of the expected transmittance so that for each group $g$ the PDTE is reduced to a transmittance interval PDTE$(g)$ for which the contribution of the fading is less detrimental. The CV-QKD protocol can be performed independently for each of these groups to obtain a secret key rate per symbol $K_{\mathrm{fad}}(\mathrm{PDTE}(g))$ and an aggregated secret key rate per symbol of
\begin{equation}
    \mathrm{K}_{\mathrm{agg}}=\sum_g P (\tau_b \in \mathrm{PDTE}(g)) K_{\mathrm{fad}}(\mathrm{PDTE}(g)).
\end{equation}

 A similar idea has been proposed in Ref.~\cite{Ruppert2019}, however here we propose to use a beacon signal to estimate the instantaneous channel transmission efficiency, instead of relying on the quantum data. This allows for a more precise estimation, also for a fast fading process. The classical beacon does not transport information related to the quantum signal and $K_{\mathrm{fad}}(\mathrm{PDTE}(g))$ is obtained using only the quantum symbols. For this reason, if the signal is tampered with in order to falsify the group classification (alter the correlation between $\tau_k$ and $\tau_b$) only a denial of service would be experienced, since the secret key rate would be reduced, as the manipulated group would suffer higher fading and more excess noise would be estimated.

In order to reduce the effect of fading, narrow PDTE intervals are desirable, but this can magnify finite size effects, since the number of symbols per group will be reduced. This compromise between PDTE interval width and number of symbols per group can be taken into account in order to optimize the division of the PDTE so that  $\mathrm{K}_{\mathrm{agg}}$ is maximal for a given PDTE and orbit duration. Technical restrictions such as the resolution available for determining $\tau_b$ can also play a role in the ideal division of the PDTE in groups.

\begin{figure}
\centering
\includegraphics[width=0.5\textwidth]{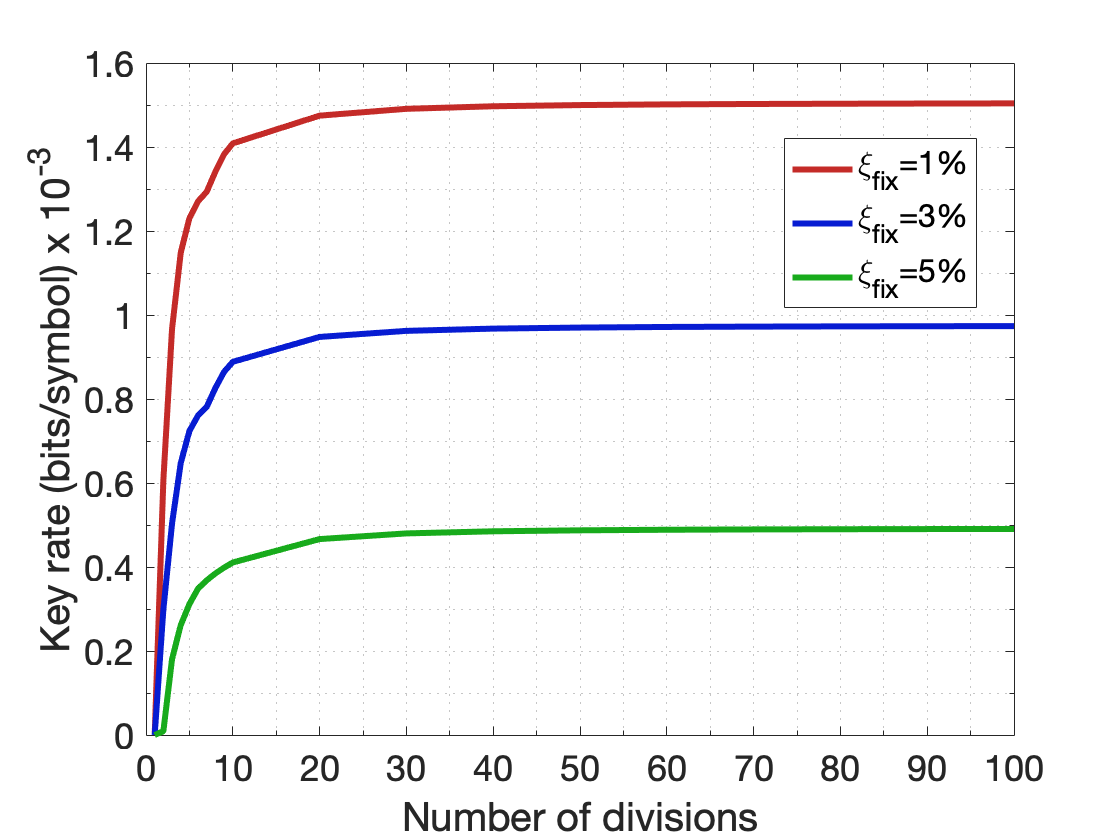}
\caption{Secret key rate for channel subdivision from 1 to 100 equally-spaced intervals for a 400 km altitude satellite. The fixed excess noise, $\xi_{\mathrm{fix}}$,  is, in S.N.U, 1 \% (red), 3\% (blue) and 5\% (green) respectively.}
\label{Scan_fixed_altitude}
\end{figure}

\begin{figure}
\centering
\includegraphics[width=0.5\textwidth]{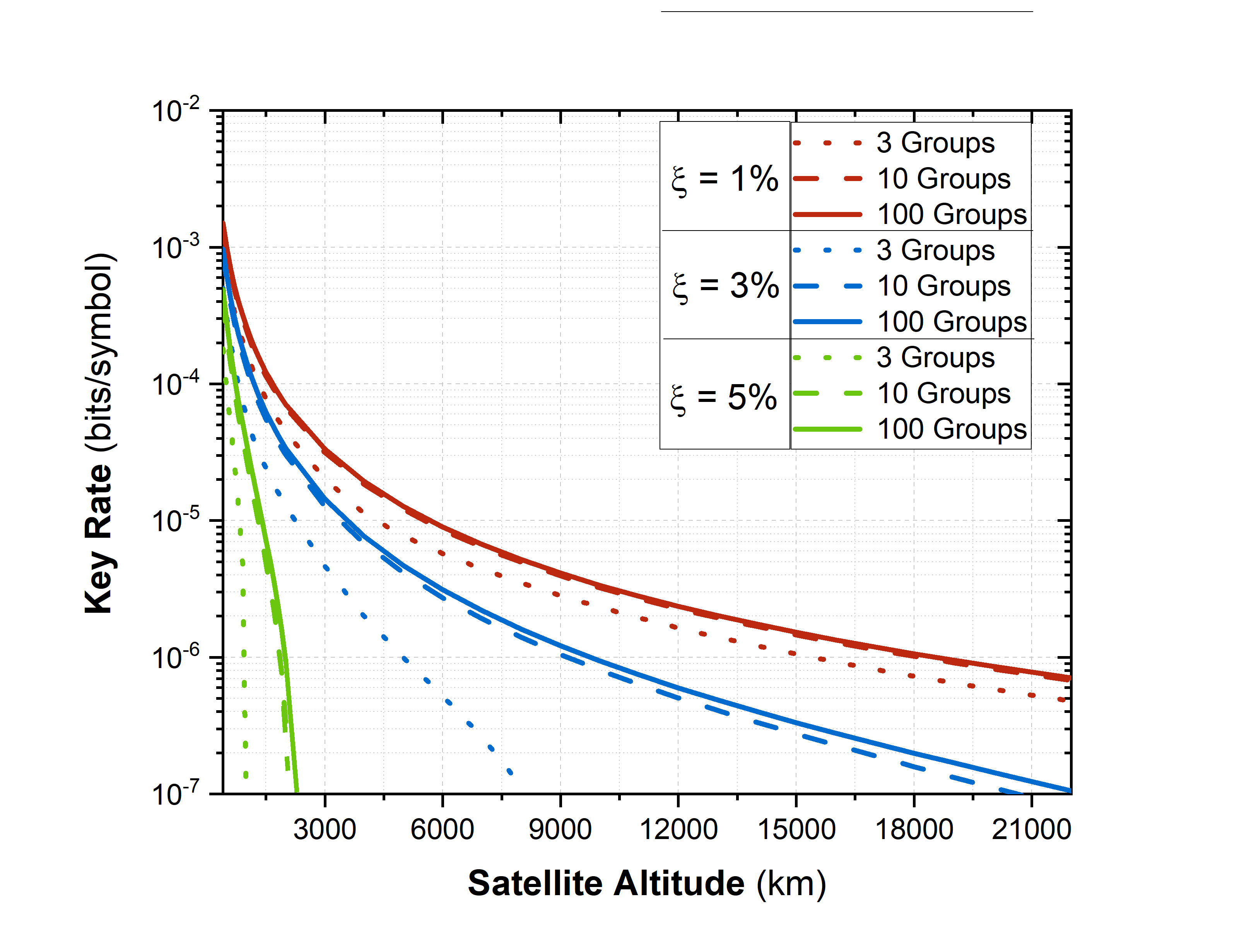}
\caption{Secret key rate for channel subdivision in 3, 10 and 100 groups and different values of the fixed excess noise, $\xi_{\mathrm{fix}}$: (in S.N.U), 1\% (red), 3\% (blue) and 5\% (green). The key rate in bits/s can be calculated by multiplying by the transmission symbol rate.}
\label{Scan_binning_dual}
\end{figure}

In our analysis we have chosen a uniform division of the PDTE and we do not treat the problem of the PDTE division optimization. We divided the whole range of transmission values in equally spaced intervals, going from a single group (corresponding to analyzing the data all together) to 100 intervals (\emph{i.e.}, close to the asymptotic limit). The results are reported in Fig.~\ref{Scan_fixed_altitude} for a satellite at $400$ km and for three values of fixed excess noise. We note that without channel subdivision no key would be possible for a 400 km orbit. To analyze the effect of the channel subdivision for all the orbits, we selected subdivisions of 3, 10 and 100 intervals for all the satellite altitudes. As shown in Fig. \ref{Scan_binning_dual}, the division of the channel transmission efficiency in 10 groups gives a total rate close to the asymptotic limit for all satellite altitudes. We underline that for this simulation the same values of beam divergence and pointing error have been used in all cases, to emphasize the impact of the orbit altitude on the key generation rate. However, due to the different satellite size and environmental disturbance, MEO satellites could in general reach better performance in terms of beam quality.\\

\noindent\textbf{Finite size analysis.} 
We complete our analysis by considering the issue of finite size effects on the estimation of parameters. It is worth noting that in satellite communication the maximum amount of time for a transmission is given by the orbital parameters and can range from few minutes to hours, depending on the satellite altitude. Moreover, as discussed previously an optimization is required if we consider the subdivision of the channel transmission efficiency for reducing the fading noise. A denser subdivision will decrease the fading noise, but will result in less populated groups, thus making the finite size effects more detrimental.

Here, we consider the uncertainty of the parameter estimation due to the limited statistics. As described in Ref.~\cite{Leverrier2010}, it is possible to account for this effect by considering a lower bound on the transmission coefficient $T=\sqrt{\tau}$ and an upper bound of the parameter $\sigma^2=1+\tau\xi$:
\begin{align}
    T_{\text{min}}&\simeq\sqrt{\tau} - z_{\epsilon_{PE}/2}\sqrt{\frac{1+\tau\xi}{m V_A}} \\
    \sigma^2_{\text{max}}&\simeq1+\tau\xi+z_{\epsilon_{PE}/2} \frac{(1+\tau\xi)\sqrt{2}}{\sqrt{m}},
\end{align}
where $m$ is the number of symbols used for parameter estimation and $z_{\epsilon_{PE}/2}$ is a parameter related to the failing probability of the parameter estimation $\epsilon_{PE}$. Here we consider $\epsilon_{PE}=10^{-10}$, which gives $z_{\epsilon_{PE}/2}= \sqrt{2} \; \textrm{erf}^{-1}(1-\epsilon_{PE}) =6.5$, where $\textrm{erf}^{-1}$ is the inverse error function. We consider the situation in which half of the symbols are used for parameter estimation and the orbit is divided in 10 intervals. This choice is not optimized and should be tailored to a specific experimental setup, however such an optimization is beyond the scope of this work.

\begin{figure}
\centering
\includegraphics[width=0.5\textwidth]{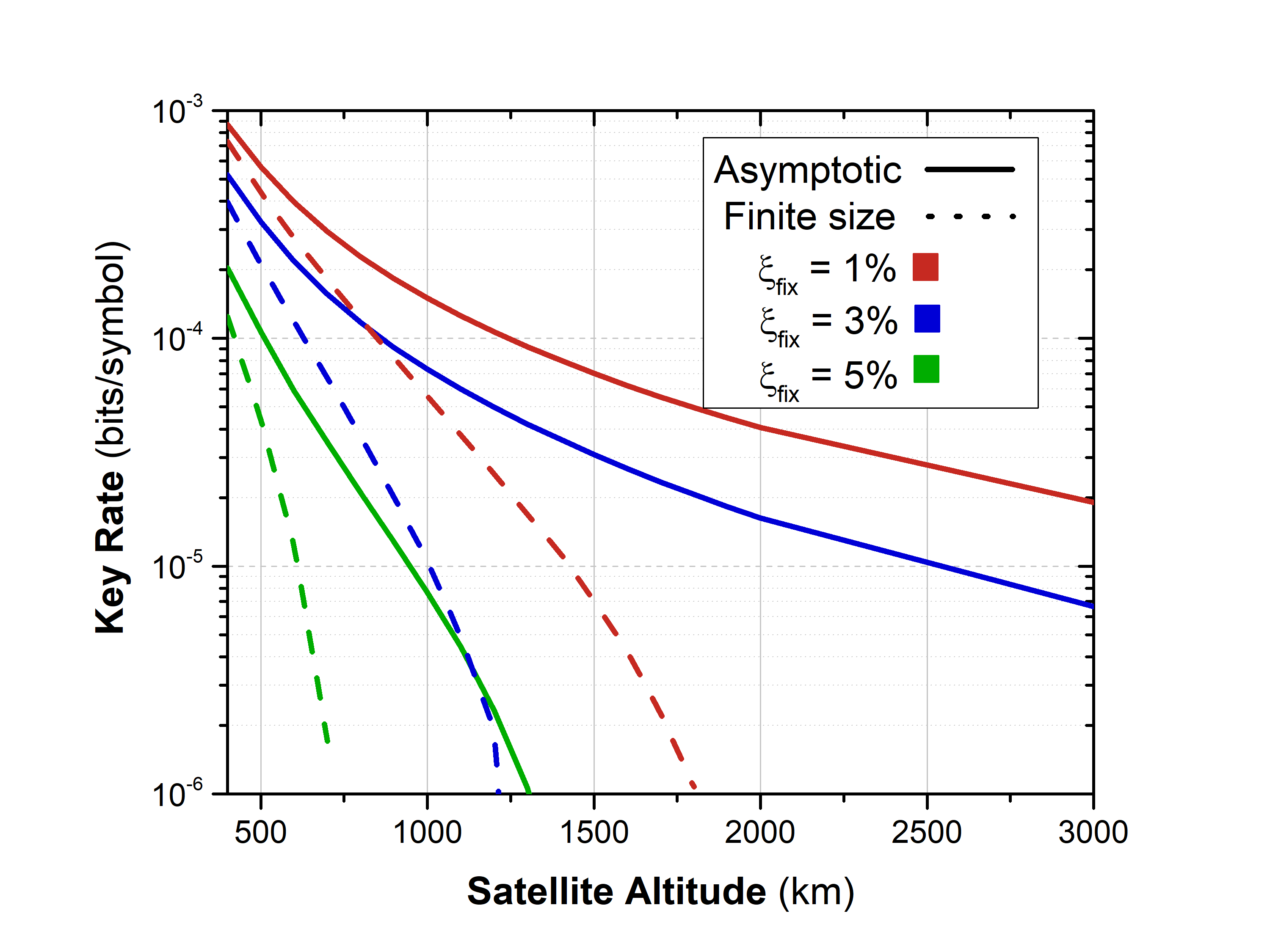}
\caption{Comparison of the key rate for LEO between the asymptotic regime (solid line) and considering finite size effects (dashed line), which have been calculated for a symbol rate of 1 Gsymbol/s. The fixed excess noise, $\xi_{\mathrm{fix}}$, is 1\% (red), 3\% (blue) 5\% (green).}
\label{finite_size}
\end{figure}

The results for the given parameters are shown in Fig.~\ref{finite_size} and highlight how the finite size effects have a remarkable impact on higher orbits, effectively precluding CV-QKD operation beyond $2000$ km when the key distillation is performed on a single satellite pass. For lower orbits, below $800$ km, the effect is only limited to a drop in the key rate. The finite size effects could be reduced increasing the transmission rate and optimizing the orbit subdivision, as well as accumulating multiple satellite passes.\\


\noindent {\large \textbf{Discussion}}\\
In this work we analyzed the feasibility of CV-QKD from satellite to a ground station. By modeling the transmission channel along a complete circular orbit, it has been possible to obtain the probability distribution of the transmission efficiency (PDTE) of the quantum channel, from which we derived the secret key generation rate both in the asymptotic case and when finite size effects are considered in the parameter estimation. To cope with channel fluctuations, typical of the satellite signal transmission, we proposed a method of data analysis based on orbit subdivision and proved its effectiveness in improving secret key generation.
The analysis provides an estimate of the expected key rate of satellite-to-ground CV-QKD and allows to constraint the experimental parameters for its realization. The obtained results show that coherent state modulation and detection is a viable option for quantum communication with LEO satellites.
The communication with higher orbits, achievable in the asymptotic limit, can be affected by finite size effects if the transmission rate is low or the orbit subdivision is not optimized. We note however that by merging multiple satellite passes, or with the implementation of higher repetition rate systems, it would be possible to extend the communication range beyond $2000$ km.
Further work is required for the comparison of the key rates achievable with continuous and discrete variable encodings in different communication scenarios.\\


\noindent {\large \textbf{Methods}}\\
\noindent \textbf{Parameter analysis.} 
Here we analyze the dependence of the secret key rate on several parameters, to obtain a better insight into which parameters affect the most the overall performance. To reduce the complexity of this multiparameter analysis, we consider here the key rate that can be obtained if the instantaneous value of the transmission efficiency is known. This case occurs when a sufficient number of symbols is exchanged within the timescale of the channel fluctuation (typically of the order of few ms) and it upper bounds the rate given by Eq.~\eqref{key_rate_fading}.
Such a situation is unrealistic in practice, however it will give us a reference for estimating the efficiency of the realistic scenario.

In this scenario, the key rate can be calculated as a weighted average, considering as weight the PDTE calculated from our channel model analysis:
\begin{equation}
    K_{UB} = \braket{\min(0, \beta I_{AB}(\tau)-\chi_{BE}(\tau))}_{\tau}.
\end{equation}

The parameters will be changed one by one, keeping the others to their reference values, expressed in Table~\ref{tab:ref_values}. The color code reflects the value of the fixed excess noise and is the same used in the main text: red, blue and green for $\xi_{\mathrm{fix}}=1, 3, 5\%$ (in S.N.U.), respectively.

\begin{figure}
\centering
\includegraphics[width=0.45\textwidth]{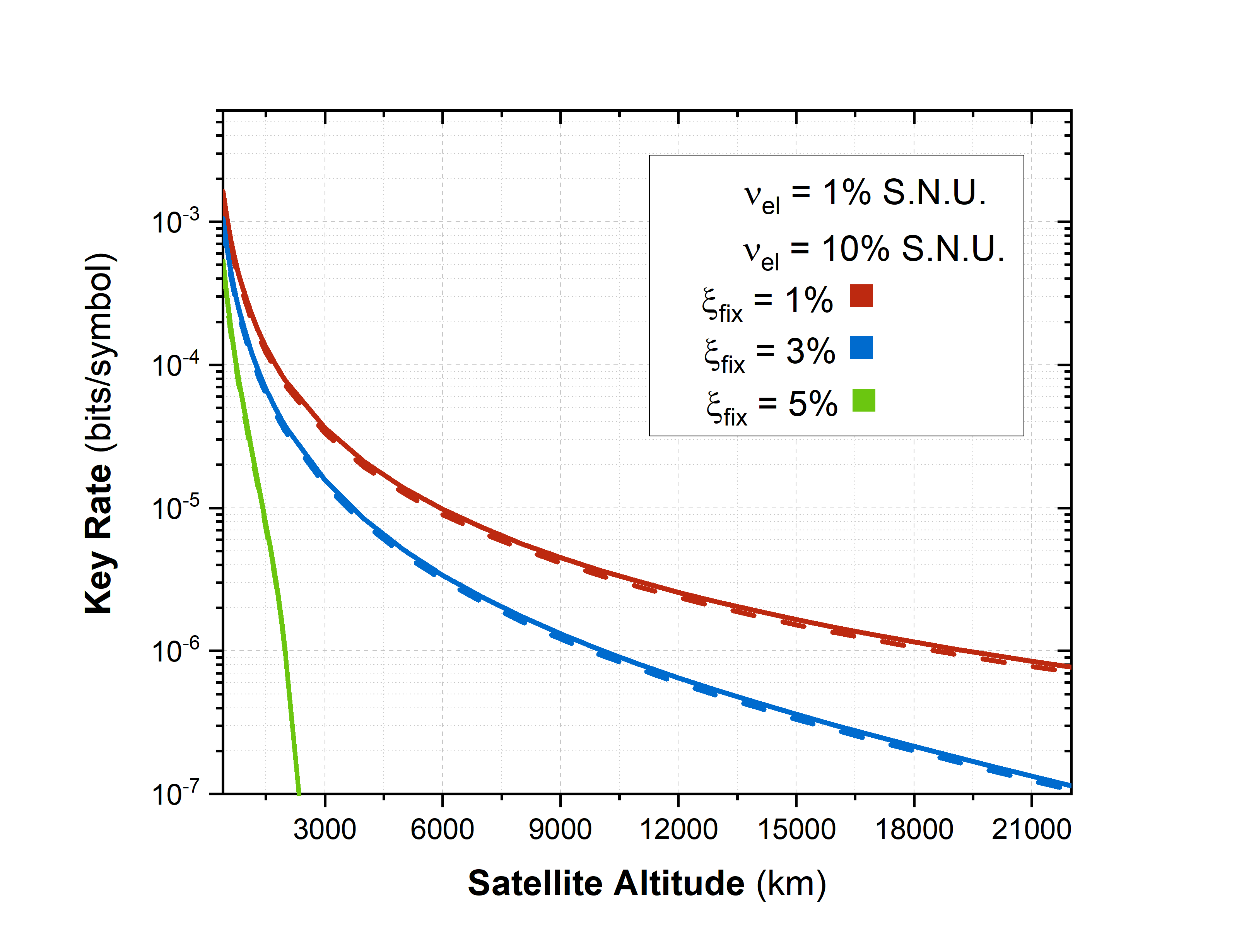}
\caption{Comparison of secret key rate for two different electronic noise $\nu_{\mathrm{el}}$ and three different excess noise values.}
\label{nu_el}
\end{figure}

In Fig.~\ref{nu_el} we vary the electronic noise of the detectors from 0.01 to 0.1 S.N.U. We notice that even with one order of magnitude increase in noise, the key rate is almost unaffected for all cases. This is mainly due to the fact that in this analysis we consider the so called ``trusted'' or ``calibrated'' scenario, in which the electronic noise is known to Bob via a constant calibration and cannot be exploited by Eve.

\begin{figure}
\centering
\includegraphics[width=0.45\textwidth]{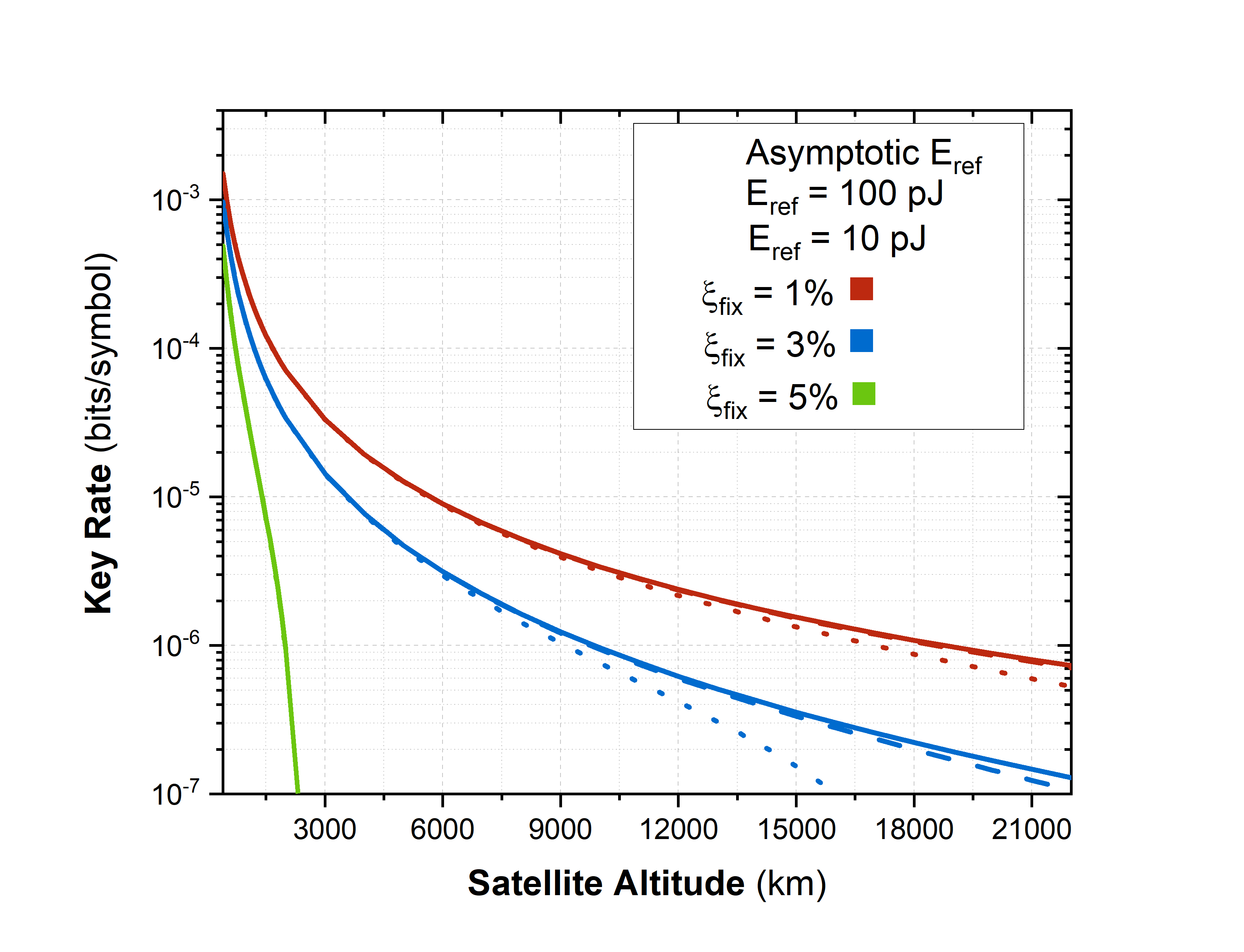}
\caption{Comparison of secret key rate for different values of the reference symbol energy $E_{\mathrm{ref}}$, for the three different excess noise values considered along the paper.}
\label{p_las}
\end{figure}

The second effect considered is the energy of the reference symbols used for phase recovery. We will illustrate the problem considering a simple phase estimation scheme operating at 1 Gsymbol/s with alternating signal and reference symbols. The time between two such symbols, $\Delta t =$ 1 ns, gives rise to a noise contribution $\xi_{\mathrm{t}}=V_A 2\pi\Delta t\Delta f$, where $\Delta f\simeq\frac{1}{\pi\tau_c}=10$ kHz is the linewidth of the two lasers and $\tau_c$ their coherence time (assumed equal for Alice and Bob). On the other hand, the phase measurement is effected by shot noise, introducing a noise  of  $\xi_{\mathrm{sn}}= \frac{V_A}{2\eta n_{\mathrm{ref}}}$, where $n_{\mathrm{ref}}=\frac{E_{\mathrm{ref}}\tau} {E_{\mathrm{photon}}}$ is the total number of photons collected, $E_{\mathrm{ref}}$ is the energy of the reference symbols and $E_{\mathrm{photon}}$ is the photon energy. The effects for different reference symbol energies is shown in Fig.~\ref{p_las}. While the effect for LEO satellites is negligible for energies above 10 pJ, for higher orbits stronger values of the reference are required to avoid any detrimental effect due to the phase alignment uncertainty, which might impose restrictions in the dynamic range of the modulators, since the optimal variance $V_A$ decreases as attenuation increases.

\begin{figure}
\centering
\includegraphics[width=0.45\textwidth]{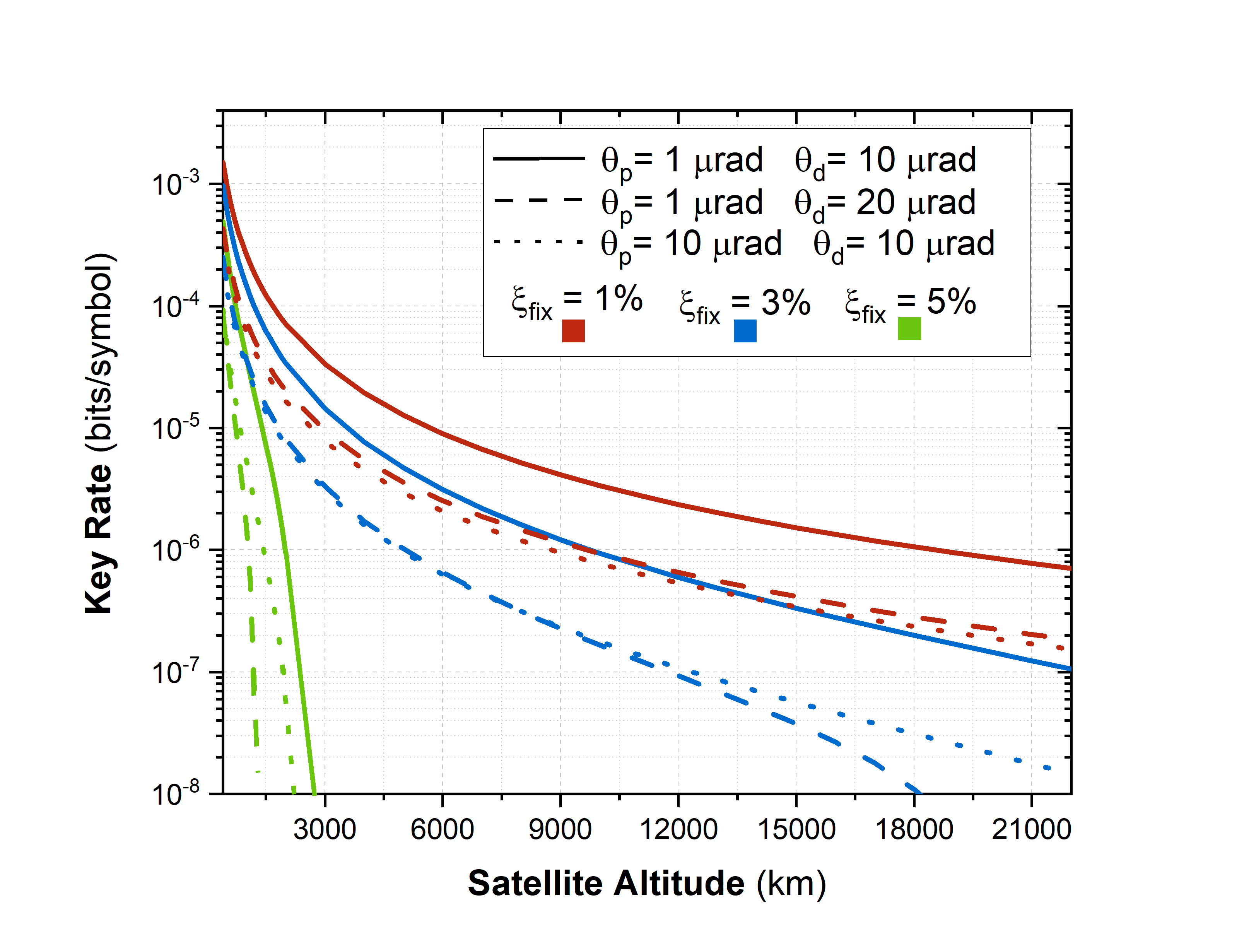}
\caption{Comparison of secret key rate for different values of pointing error, $\theta_p$, and divergence angle, $\theta_d$ at the three excess noise values.}
\label{beam_scan}
\end{figure}

Finally we consider the impact of the downlink beam characteristics, namely the pointing error and the beam divergence, on the final key rate. As expected, these values have a strong impact in all the configurations shown in Fig.~\ref{beam_scan}, underlying the importance of a high quality beam propagation for satellite CV-QKD.

\medskip

\noindent \textbf{Previous treatment of fading in the literature.} 
Reference \cite{Papanastasiou2018} considers two scenarios: slow fading where the transmission efficiency fluctuates at a slower rate than the key establishment rate, and fast fading where the transmission value fluctuates significantly during a single key extraction procedure. The second scenario is similar to ours, but the expression of the authors for the secret key rate differs since they obtain
\begin{align}
K_{\text{fast fading}} = \beta I_AB^{\eta_{\min}} - \int d\tau P_\tau \chi(E;y),
\end{align}
where the transmission efficiency $\tau=T^2$ is uniformly distributed with distribution $P_\tau$ over some interval $[\tau_{\min}, \tau_{\max}]$.
In other words, they take the most pessimistic value of $I_{AB}$ (corresponding to the lowest transmission value) and consider the average of the Holevo information between Eve and the raw key, over the possible fading values.

In contrast, we agree with the estimate for the Holevo information from Ref.~\cite{Usenko2012} but take a more conservative value for the mutual information $I_{AB}$ since their value is computed for a Gaussian modulation that would yield the same covariance matrix. We have instead argued that one needs to carefully consider the classical channel mapping $Y$ to $X$ (in the reverse reconciliation procedure). This is a fading channel where one can take advantage of the pilot signals to get a rough estimate of the fading coefficient. 
This implies that one can approximate the capacity of that channel with the average of the capacities of an AWGN channel over the value of the fading parameter.\\ 



\noindent {\large \textbf{Acknowledgments}}\\
We thank Michel Maignan, Mathias Van den Bossche, Philippe Grangier and Matteo Schiavon for helpful discussions. We acknowledge support from the European Commission's Horizon 2020 Research and Innovation Programme under Grant Agreement No. 820466 (CiViQ) and from the European COST Action MP1403 ``Nanoscale Quantum Optics''. VRR acknowledges support from the DGA. DD acknowledges support from the Italian Space Agency.\\



\bibliographystyle{apsrev4-1}
\bibliography{main}

\begin{thebibliography}{69}%
\makeatletter
\providecommand \@ifxundefined [1]{%
 \@ifx{#1\undefined}
}%
\providecommand \@ifnum [1]{%
 \ifnum #1\expandafter \@firstoftwo
 \else \expandafter \@secondoftwo
 \fi
}%
\providecommand \@ifx [1]{%
 \ifx #1\expandafter \@firstoftwo
 \else \expandafter \@secondoftwo
 \fi
}%
\providecommand \natexlab [1]{#1}%
\providecommand \enquote  [1]{``#1''}%
\providecommand \bibnamefont  [1]{#1}%
\providecommand \bibfnamefont [1]{#1}%
\providecommand \citenamefont [1]{#1}%
\providecommand \href@noop [0]{\@secondoftwo}%
\providecommand \href [0]{\begingroup \@sanitize@url \@href}%
\providecommand \@href[1]{\@@startlink{#1}\@@href}%
\providecommand \@@href[1]{\endgroup#1\@@endlink}%
\providecommand \@sanitize@url [0]{\catcode `\\12\catcode `\$12\catcode
  `\&12\catcode `\#12\catcode `\^12\catcode `\_12\catcode `\%12\relax}%
\providecommand \@@startlink[1]{}%
\providecommand \@@endlink[0]{}%
\providecommand \url  [0]{\begingroup\@sanitize@url \@url }%
\providecommand \@url [1]{\endgroup\@href {#1}{\urlprefix }}%
\providecommand \urlprefix  [0]{URL }%
\providecommand \Eprint [0]{\href }%
\providecommand \doibase [0]{http://dx.doi.org/}%
\providecommand \selectlanguage [0]{\@gobble}%
\providecommand \bibinfo  [0]{\@secondoftwo}%
\providecommand \bibfield  [0]{\@secondoftwo}%
\providecommand \translation [1]{[#1]}%
\providecommand \BibitemOpen [0]{}%
\providecommand \bibitemStop [0]{}%
\providecommand \bibitemNoStop [0]{.\EOS\space}%
\providecommand \EOS [0]{\spacefactor3000\relax}%
\providecommand \BibitemShut  [1]{\csname bibitem#1\endcsname}%
\let\auto@bib@innerbib\@empty
\bibitem [{\citenamefont {Scarani}\ \emph {et~al.}(2009)\citenamefont
  {Scarani}, \citenamefont {Bechmann-Pasquinucci}, \citenamefont {Cerf},
  \citenamefont {Du{\v{s}}ek}, \citenamefont {L{\"{u}}tkenhaus},\ and\
  \citenamefont {Peev}}]{Scarani2009}%
  \BibitemOpen
  \bibfield  {author} {\bibinfo {author} {\bibfnamefont {V.}~\bibnamefont
  {Scarani}}, \bibinfo {author} {\bibfnamefont {H.}~\bibnamefont
  {Bechmann-Pasquinucci}}, \bibinfo {author} {\bibfnamefont {N.~J.}\
  \bibnamefont {Cerf}}, \bibinfo {author} {\bibfnamefont {M.}~\bibnamefont
  {Du{\v{s}}ek}}, \bibinfo {author} {\bibfnamefont {N.}~\bibnamefont
  {L{\"{u}}tkenhaus}}, \ and\ \bibinfo {author} {\bibfnamefont
  {M.}~\bibnamefont {Peev}},\ }\href {\doibase 10.1103/RevModPhys.81.1301}
  {\bibfield  {journal} {\bibinfo  {journal} {Rev. Mod. Phys.}\ }\textbf
  {\bibinfo {volume} {81}},\ \bibinfo {pages} {1301} (\bibinfo {year}
  {2009})}\BibitemShut {NoStop}%
\bibitem [{\citenamefont {Diamanti}\ \emph {et~al.}(2016)\citenamefont
  {Diamanti}, \citenamefont {Lo}, \citenamefont {Qi},\ and\ \citenamefont
  {Yuan}}]{Diamanti16}%
  \BibitemOpen
  \bibfield  {author} {\bibinfo {author} {\bibfnamefont {E.}~\bibnamefont
  {Diamanti}}, \bibinfo {author} {\bibfnamefont {H.-K.}\ \bibnamefont {Lo}},
  \bibinfo {author} {\bibfnamefont {B.}~\bibnamefont {Qi}}, \ and\ \bibinfo
  {author} {\bibfnamefont {Z.}~\bibnamefont {Yuan}},\ }\href@noop {} {\bibfield
   {journal} {\bibinfo  {journal} {npj Quantum Information}\ }\textbf {\bibinfo
  {volume} {2}},\ \bibinfo {pages} {16025} (\bibinfo {year}
  {2016})}\BibitemShut {NoStop}%
\bibitem [{\citenamefont {Jouguet}\ \emph {et~al.}(2013)\citenamefont
  {Jouguet}, \citenamefont {Kunz-Jacques}, \citenamefont {Leverrier},
  \citenamefont {Grangier},\ and\ \citenamefont {Diamanti}}]{Jouguet2013a}%
  \BibitemOpen
  \bibfield  {author} {\bibinfo {author} {\bibfnamefont {P.}~\bibnamefont
  {Jouguet}}, \bibinfo {author} {\bibfnamefont {S.}~\bibnamefont
  {Kunz-Jacques}}, \bibinfo {author} {\bibfnamefont {A.}~\bibnamefont
  {Leverrier}}, \bibinfo {author} {\bibfnamefont {P.}~\bibnamefont {Grangier}},
  \ and\ \bibinfo {author} {\bibfnamefont {E.}~\bibnamefont {Diamanti}},\
  }\href {\doibase 10.1038/nphoton.2013.63} {\bibfield  {journal} {\bibinfo
  {journal} {Nature Photonics}\ }\textbf {\bibinfo {volume} {7}},\ \bibinfo
  {pages} {378} (\bibinfo {year} {2013})}\BibitemShut {NoStop}%
\bibitem [{\citenamefont {Tang}\ \emph {et~al.}(2014)\citenamefont {Tang},
  \citenamefont {Yin}, \citenamefont {Chen}, \citenamefont {Liu}, \citenamefont
  {Zhang}, \citenamefont {Jiang}, \citenamefont {Zhang}, \citenamefont {Wang},
  \citenamefont {You}, \citenamefont {Guan}, \citenamefont {Yang},
  \citenamefont {Wang}, \citenamefont {Liang}, \citenamefont {Zhang},
  \citenamefont {Zhou}, \citenamefont {Ma}, \citenamefont {Chen}, \citenamefont
  {Zhang},\ and\ \citenamefont {Pan}}]{Tang2014}%
  \BibitemOpen
  \bibfield  {author} {\bibinfo {author} {\bibfnamefont {Y.~L.}\ \bibnamefont
  {Tang}}, \bibinfo {author} {\bibfnamefont {H.~L.}\ \bibnamefont {Yin}},
  \bibinfo {author} {\bibfnamefont {S.~J.}\ \bibnamefont {Chen}}, \bibinfo
  {author} {\bibfnamefont {Y.}~\bibnamefont {Liu}}, \bibinfo {author}
  {\bibfnamefont {W.~J.}\ \bibnamefont {Zhang}}, \bibinfo {author}
  {\bibfnamefont {X.}~\bibnamefont {Jiang}}, \bibinfo {author} {\bibfnamefont
  {L.}~\bibnamefont {Zhang}}, \bibinfo {author} {\bibfnamefont
  {J.}~\bibnamefont {Wang}}, \bibinfo {author} {\bibfnamefont {L.~X.}\
  \bibnamefont {You}}, \bibinfo {author} {\bibfnamefont {J.~Y.}\ \bibnamefont
  {Guan}}, \bibinfo {author} {\bibfnamefont {D.~X.}\ \bibnamefont {Yang}},
  \bibinfo {author} {\bibfnamefont {Z.}~\bibnamefont {Wang}}, \bibinfo {author}
  {\bibfnamefont {H.}~\bibnamefont {Liang}}, \bibinfo {author} {\bibfnamefont
  {Z.}~\bibnamefont {Zhang}}, \bibinfo {author} {\bibfnamefont
  {N.}~\bibnamefont {Zhou}}, \bibinfo {author} {\bibfnamefont {X.}~\bibnamefont
  {Ma}}, \bibinfo {author} {\bibfnamefont {T.~Y.}\ \bibnamefont {Chen}},
  \bibinfo {author} {\bibfnamefont {Q.}~\bibnamefont {Zhang}}, \ and\ \bibinfo
  {author} {\bibfnamefont {J.~W.}\ \bibnamefont {Pan}},\ }\href {\doibase
  10.1103/PhysRevLett.113.190501} {\bibfield  {journal} {\bibinfo  {journal}
  {Phys. Rev. Lett.}\ }\textbf {\bibinfo {volume} {113}},\ \bibinfo {pages} {1}
  (\bibinfo {year} {2014})}\BibitemShut {NoStop}%
\bibitem [{\citenamefont {Yin}\ \emph {et~al.}(2016)\citenamefont {Yin},
  \citenamefont {Chen}, \citenamefont {Yu}, \citenamefont {Liu}, \citenamefont
  {You}, \citenamefont {Zhou}, \citenamefont {Chen}, \citenamefont {Mao},
  \citenamefont {Huang}, \citenamefont {Zhang}, \citenamefont {Chen},
  \citenamefont {Li}, \citenamefont {Nolan}, \citenamefont {Zhou},
  \citenamefont {Jiang}, \citenamefont {Wang}, \citenamefont {Zhang},
  \citenamefont {Wang},\ and\ \citenamefont {Pan}}]{Yin2016}%
  \BibitemOpen
  \bibfield  {author} {\bibinfo {author} {\bibfnamefont {H.~L.}\ \bibnamefont
  {Yin}}, \bibinfo {author} {\bibfnamefont {T.~Y.}\ \bibnamefont {Chen}},
  \bibinfo {author} {\bibfnamefont {Z.~W.}\ \bibnamefont {Yu}}, \bibinfo
  {author} {\bibfnamefont {H.}~\bibnamefont {Liu}}, \bibinfo {author}
  {\bibfnamefont {L.~X.}\ \bibnamefont {You}}, \bibinfo {author} {\bibfnamefont
  {Y.~H.}\ \bibnamefont {Zhou}}, \bibinfo {author} {\bibfnamefont {S.~J.}\
  \bibnamefont {Chen}}, \bibinfo {author} {\bibfnamefont {Y.}~\bibnamefont
  {Mao}}, \bibinfo {author} {\bibfnamefont {M.~Q.}\ \bibnamefont {Huang}},
  \bibinfo {author} {\bibfnamefont {W.~J.}\ \bibnamefont {Zhang}}, \bibinfo
  {author} {\bibfnamefont {H.}~\bibnamefont {Chen}}, \bibinfo {author}
  {\bibfnamefont {M.~J.}\ \bibnamefont {Li}}, \bibinfo {author} {\bibfnamefont
  {D.}~\bibnamefont {Nolan}}, \bibinfo {author} {\bibfnamefont
  {F.}~\bibnamefont {Zhou}}, \bibinfo {author} {\bibfnamefont {X.}~\bibnamefont
  {Jiang}}, \bibinfo {author} {\bibfnamefont {Z.}~\bibnamefont {Wang}},
  \bibinfo {author} {\bibfnamefont {Q.}~\bibnamefont {Zhang}}, \bibinfo
  {author} {\bibfnamefont {X.~B.}\ \bibnamefont {Wang}}, \ and\ \bibinfo
  {author} {\bibfnamefont {J.~W.}\ \bibnamefont {Pan}},\ }\href {\doibase
  10.1103/PhysRevLett.117.190501} {\bibfield  {journal} {\bibinfo  {journal}
  {Phys. Rev. Lett.}\ }\textbf {\bibinfo {volume} {117}},\ \bibinfo {pages}
  {190501} (\bibinfo {year} {2016})}\BibitemShut {NoStop}%
\bibitem [{\citenamefont {Boaron}\ \emph {et~al.}(2018)\citenamefont {Boaron},
  \citenamefont {Boso}, \citenamefont {Rusca}, \citenamefont {Vulliez},
  \citenamefont {Autebert}, \citenamefont {Caloz}, \citenamefont {Perrenoud},
  \citenamefont {Gras}, \citenamefont {Bussi{\`{e}}res}, \citenamefont {Li},
  \citenamefont {Nolan}, \citenamefont {Martin},\ and\ \citenamefont
  {Zbinden}}]{Boaron2018}%
  \BibitemOpen
  \bibfield  {author} {\bibinfo {author} {\bibfnamefont {A.}~\bibnamefont
  {Boaron}}, \bibinfo {author} {\bibfnamefont {G.}~\bibnamefont {Boso}},
  \bibinfo {author} {\bibfnamefont {D.}~\bibnamefont {Rusca}}, \bibinfo
  {author} {\bibfnamefont {C.}~\bibnamefont {Vulliez}}, \bibinfo {author}
  {\bibfnamefont {C.}~\bibnamefont {Autebert}}, \bibinfo {author}
  {\bibfnamefont {M.}~\bibnamefont {Caloz}}, \bibinfo {author} {\bibfnamefont
  {M.}~\bibnamefont {Perrenoud}}, \bibinfo {author} {\bibfnamefont
  {G.}~\bibnamefont {Gras}}, \bibinfo {author} {\bibfnamefont {F.}~\bibnamefont
  {Bussi{\`{e}}res}}, \bibinfo {author} {\bibfnamefont {M.~J.}\ \bibnamefont
  {Li}}, \bibinfo {author} {\bibfnamefont {D.}~\bibnamefont {Nolan}}, \bibinfo
  {author} {\bibfnamefont {A.}~\bibnamefont {Martin}}, \ and\ \bibinfo {author}
  {\bibfnamefont {H.}~\bibnamefont {Zbinden}},\ }\href {\doibase
  10.1103/PhysRevLett.121.190502} {\bibfield  {journal} {\bibinfo  {journal}
  {Phys. Rev. Lett.}\ }\textbf {\bibinfo {volume} {121}},\ \bibinfo {pages}
  {190502} (\bibinfo {year} {2018})}\BibitemShut {NoStop}%
\bibitem [{\citenamefont {Pirandola}\ \emph {et~al.}(2017)\citenamefont
  {Pirandola}, \citenamefont {Laurenza}, \citenamefont {Ottaviani},\ and\
  \citenamefont {Banchi}}]{Pirandola2017}%
  \BibitemOpen
  \bibfield  {author} {\bibinfo {author} {\bibfnamefont {S.}~\bibnamefont
  {Pirandola}}, \bibinfo {author} {\bibfnamefont {R.}~\bibnamefont {Laurenza}},
  \bibinfo {author} {\bibfnamefont {C.}~\bibnamefont {Ottaviani}}, \ and\
  \bibinfo {author} {\bibfnamefont {L.}~\bibnamefont {Banchi}},\ }\href
  {\doibase 10.1038/ncomms15043} {\bibfield  {journal} {\bibinfo  {journal}
  {Nature Communications}\ }\textbf {\bibinfo {volume} {8}},\ \bibinfo {pages}
  {15043} (\bibinfo {year} {2017})}\BibitemShut {NoStop}%
\bibitem [{\citenamefont {Briegel}\ \emph {et~al.}(1998)\citenamefont
  {Briegel}, \citenamefont {D\"{u}r}, \citenamefont {Cirac},\ and\
  \citenamefont {Zoller}}]{Briegel1998}%
  \BibitemOpen
  \bibfield  {author} {\bibinfo {author} {\bibfnamefont {H.-J.}\ \bibnamefont
  {Briegel}}, \bibinfo {author} {\bibfnamefont {W.}~\bibnamefont {D\"{u}r}},
  \bibinfo {author} {\bibfnamefont {J.~I.}\ \bibnamefont {Cirac}}, \ and\
  \bibinfo {author} {\bibfnamefont {P.}~\bibnamefont {Zoller}},\ }\href
  {\doibase 10.1103/physrevlett.81.5932} {\bibfield  {journal} {\bibinfo
  {journal} {Phys. Rev. Lett.}\ }\textbf {\bibinfo {volume} {81}},\ \bibinfo
  {pages} {5932} (\bibinfo {year} {1998})}\BibitemShut {NoStop}%
\bibitem [{\citenamefont {Jiang}\ \emph {et~al.}(2009)\citenamefont {Jiang},
  \citenamefont {Taylor}, \citenamefont {Nemoto}, \citenamefont {Munro},
  \citenamefont {Meter},\ and\ \citenamefont {Lukin}}]{Jiang2009}%
  \BibitemOpen
  \bibfield  {author} {\bibinfo {author} {\bibfnamefont {L.}~\bibnamefont
  {Jiang}}, \bibinfo {author} {\bibfnamefont {J.~M.}\ \bibnamefont {Taylor}},
  \bibinfo {author} {\bibfnamefont {K.}~\bibnamefont {Nemoto}}, \bibinfo
  {author} {\bibfnamefont {W.~J.}\ \bibnamefont {Munro}}, \bibinfo {author}
  {\bibfnamefont {R.~V.}\ \bibnamefont {Meter}}, \ and\ \bibinfo {author}
  {\bibfnamefont {M.~D.}\ \bibnamefont {Lukin}},\ }\href {\doibase
  10.1103/physreva.79.032325} {\bibfield  {journal} {\bibinfo  {journal} {Phys.
  Rev. A}\ }\textbf {\bibinfo {volume} {79}},\ \bibinfo {pages} {032325}
  (\bibinfo {year} {2009})}\BibitemShut {NoStop}%
\bibitem [{\citenamefont {Sangouard}\ \emph {et~al.}(2011)\citenamefont
  {Sangouard}, \citenamefont {Simon}, \citenamefont {de~Riedmatten},\ and\
  \citenamefont {Gisin}}]{Sangouard2011}%
  \BibitemOpen
  \bibfield  {author} {\bibinfo {author} {\bibfnamefont {N.}~\bibnamefont
  {Sangouard}}, \bibinfo {author} {\bibfnamefont {C.}~\bibnamefont {Simon}},
  \bibinfo {author} {\bibfnamefont {H.}~\bibnamefont {de~Riedmatten}}, \ and\
  \bibinfo {author} {\bibfnamefont {N.}~\bibnamefont {Gisin}},\ }\href
  {\doibase 10.1103/revmodphys.83.33} {\bibfield  {journal} {\bibinfo
  {journal} {Rev. Mod. Phys.}\ }\textbf {\bibinfo {volume} {83}},\ \bibinfo
  {pages} {33} (\bibinfo {year} {2011})}\BibitemShut {NoStop}%
\bibitem [{\citenamefont {Azuma}\ \emph {et~al.}(2015)\citenamefont {Azuma},
  \citenamefont {Tamaki},\ and\ \citenamefont {Lo}}]{Azuma2015}%
  \BibitemOpen
  \bibfield  {author} {\bibinfo {author} {\bibfnamefont {K.}~\bibnamefont
  {Azuma}}, \bibinfo {author} {\bibfnamefont {K.}~\bibnamefont {Tamaki}}, \
  and\ \bibinfo {author} {\bibfnamefont {H.-K.}\ \bibnamefont {Lo}},\ }\href
  {\doibase 10.1038/ncomms7787} {\bibfield  {journal} {\bibinfo  {journal}
  {Nature Communications}\ }\textbf {\bibinfo {volume} {6}},\ \bibinfo {pages}
  {6787} (\bibinfo {year} {2015})}\BibitemShut {NoStop}%
\bibitem [{\citenamefont {Vinay}\ and\ \citenamefont {Kok}(2017)}]{Vinay2017}%
  \BibitemOpen
  \bibfield  {author} {\bibinfo {author} {\bibfnamefont {S.~E.}\ \bibnamefont
  {Vinay}}\ and\ \bibinfo {author} {\bibfnamefont {P.}~\bibnamefont {Kok}},\
  }\href {\doibase 10.1103/physreva.95.052336} {\bibfield  {journal} {\bibinfo
  {journal} {Phys. Rev. A}\ }\textbf {\bibinfo {volume} {95}},\ \bibinfo
  {pages} {052336} (\bibinfo {year} {2017})}\BibitemShut {NoStop}%
\bibitem [{\citenamefont {Li}\ \emph {et~al.}(2019)\citenamefont {Li},
  \citenamefont {Zhang}, \citenamefont {Yin}, \citenamefont {Liu},
  \citenamefont {Hu}, \citenamefont {Fang}, \citenamefont {Fei}, \citenamefont
  {Jiang}, \citenamefont {Zhang}, \citenamefont {Li}, \citenamefont {Liu},
  \citenamefont {Xu}, \citenamefont {Chen},\ and\ \citenamefont
  {Pan}}]{Li2019}%
  \BibitemOpen
  \bibfield  {author} {\bibinfo {author} {\bibfnamefont {Z.-D.}\ \bibnamefont
  {Li}}, \bibinfo {author} {\bibfnamefont {R.}~\bibnamefont {Zhang}}, \bibinfo
  {author} {\bibfnamefont {X.-F.}\ \bibnamefont {Yin}}, \bibinfo {author}
  {\bibfnamefont {L.-Z.}\ \bibnamefont {Liu}}, \bibinfo {author} {\bibfnamefont
  {Y.}~\bibnamefont {Hu}}, \bibinfo {author} {\bibfnamefont {Y.-Q.}\
  \bibnamefont {Fang}}, \bibinfo {author} {\bibfnamefont {Y.-Y.}\ \bibnamefont
  {Fei}}, \bibinfo {author} {\bibfnamefont {X.}~\bibnamefont {Jiang}}, \bibinfo
  {author} {\bibfnamefont {J.}~\bibnamefont {Zhang}}, \bibinfo {author}
  {\bibfnamefont {L.}~\bibnamefont {Li}}, \bibinfo {author} {\bibfnamefont
  {N.-L.}\ \bibnamefont {Liu}}, \bibinfo {author} {\bibfnamefont
  {F.}~\bibnamefont {Xu}}, \bibinfo {author} {\bibfnamefont {Y.-A.}\
  \bibnamefont {Chen}}, \ and\ \bibinfo {author} {\bibfnamefont {J.-W.}\
  \bibnamefont {Pan}},\ }\href {\doibase 10.1038/s41566-019-0468-5} {\bibfield
  {journal} {\bibinfo  {journal} {Nature Photonics}\ }\textbf {\bibinfo
  {volume} {13}},\ \bibinfo {pages} {644} (\bibinfo {year} {2019})}\BibitemShut
  {NoStop}%
\bibitem [{\citenamefont {Bhaskar}\ \emph {et~al.}(2020)\citenamefont
  {Bhaskar}, \citenamefont {Riedinger}, \citenamefont {Machielse},
  \citenamefont {Levonian}, \citenamefont {Nguyen}, \citenamefont {Knall},
  \citenamefont {Park}, \citenamefont {Englund}, \citenamefont {Loncar},
  \citenamefont {Sukachev},\ and\ \citenamefont {Lukin}}]{Bhaskar2020}%
  \BibitemOpen
  \bibfield  {author} {\bibinfo {author} {\bibfnamefont {M.~K.}\ \bibnamefont
  {Bhaskar}}, \bibinfo {author} {\bibfnamefont {R.}~\bibnamefont {Riedinger}},
  \bibinfo {author} {\bibfnamefont {B.}~\bibnamefont {Machielse}}, \bibinfo
  {author} {\bibfnamefont {D.~S.}\ \bibnamefont {Levonian}}, \bibinfo {author}
  {\bibfnamefont {C.~T.}\ \bibnamefont {Nguyen}}, \bibinfo {author}
  {\bibfnamefont {E.~N.}\ \bibnamefont {Knall}}, \bibinfo {author}
  {\bibfnamefont {H.}~\bibnamefont {Park}}, \bibinfo {author} {\bibfnamefont
  {D.}~\bibnamefont {Englund}}, \bibinfo {author} {\bibfnamefont
  {M.}~\bibnamefont {Loncar}}, \bibinfo {author} {\bibfnamefont {D.~D.}\
  \bibnamefont {Sukachev}}, \ and\ \bibinfo {author} {\bibfnamefont {M.~D.}\
  \bibnamefont {Lukin}},\ }\href {\doibase 10.1038/s41586-020-2103-5}
  {\bibfield  {journal} {\bibinfo  {journal} {Nature}\ }\textbf {\bibinfo
  {volume} {580}},\ \bibinfo {pages} {60} (\bibinfo {year} {2020})}\BibitemShut
  {NoStop}%
\bibitem [{\citenamefont {Villoresi}\ \emph {et~al.}(2008)\citenamefont
  {Villoresi}, \citenamefont {Jennewein}, \citenamefont {Tamburini},
  \citenamefont {Aspelmeyer}, \citenamefont {Bonato}, \citenamefont {Ursin},
  \citenamefont {Pernechele}, \citenamefont {Luceri}, \citenamefont {Bianco},
  \citenamefont {Zeilinger},\ and\ \citenamefont {Barbieri}}]{Villoresi2008}%
  \BibitemOpen
  \bibfield  {author} {\bibinfo {author} {\bibfnamefont {P.}~\bibnamefont
  {Villoresi}}, \bibinfo {author} {\bibfnamefont {T.}~\bibnamefont
  {Jennewein}}, \bibinfo {author} {\bibfnamefont {F.}~\bibnamefont
  {Tamburini}}, \bibinfo {author} {\bibfnamefont {M.}~\bibnamefont
  {Aspelmeyer}}, \bibinfo {author} {\bibfnamefont {C.}~\bibnamefont {Bonato}},
  \bibinfo {author} {\bibfnamefont {R.}~\bibnamefont {Ursin}}, \bibinfo
  {author} {\bibfnamefont {C.}~\bibnamefont {Pernechele}}, \bibinfo {author}
  {\bibfnamefont {V.}~\bibnamefont {Luceri}}, \bibinfo {author} {\bibfnamefont
  {G.}~\bibnamefont {Bianco}}, \bibinfo {author} {\bibfnamefont
  {A.}~\bibnamefont {Zeilinger}}, \ and\ \bibinfo {author} {\bibfnamefont
  {C.}~\bibnamefont {Barbieri}},\ }\href {\doibase
  10.1088/1367-2630/10/3/033038} {\bibfield  {journal} {\bibinfo  {journal}
  {New J. Phys.}\ }\textbf {\bibinfo {volume} {10}},\ \bibinfo {pages} {033038}
  (\bibinfo {year} {2008})}\BibitemShut {NoStop}%
\bibitem [{\citenamefont {Yin}\ \emph {et~al.}(2013)\citenamefont {Yin},
  \citenamefont {Cao}, \citenamefont {Liu}, \citenamefont {Pan}, \citenamefont
  {Wang}, \citenamefont {Yang}, \citenamefont {Zhang}, \citenamefont {Yang},
  \citenamefont {Chen}, \citenamefont {Peng},\ and\ \citenamefont
  {Pan}}]{Yin2013a}%
  \BibitemOpen
  \bibfield  {author} {\bibinfo {author} {\bibfnamefont {J.}~\bibnamefont
  {Yin}}, \bibinfo {author} {\bibfnamefont {Y.}~\bibnamefont {Cao}}, \bibinfo
  {author} {\bibfnamefont {S.-B.}\ \bibnamefont {Liu}}, \bibinfo {author}
  {\bibfnamefont {G.-S.}\ \bibnamefont {Pan}}, \bibinfo {author} {\bibfnamefont
  {J.-H.}\ \bibnamefont {Wang}}, \bibinfo {author} {\bibfnamefont
  {T.}~\bibnamefont {Yang}}, \bibinfo {author} {\bibfnamefont {Z.-P.}\
  \bibnamefont {Zhang}}, \bibinfo {author} {\bibfnamefont {F.-M.}\ \bibnamefont
  {Yang}}, \bibinfo {author} {\bibfnamefont {Y.-A.}\ \bibnamefont {Chen}},
  \bibinfo {author} {\bibfnamefont {C.-Z.}\ \bibnamefont {Peng}}, \ and\
  \bibinfo {author} {\bibfnamefont {J.-W.}\ \bibnamefont {Pan}},\ }\href
  {\doibase 10.1364/OE.21.020032} {\bibfield  {journal} {\bibinfo  {journal}
  {Opt. Express}\ }\textbf {\bibinfo {volume} {21}},\ \bibinfo {pages} {20032}
  (\bibinfo {year} {2013})}\BibitemShut {NoStop}%
\bibitem [{\citenamefont {Vallone}\ \emph {et~al.}(2015)\citenamefont
  {Vallone}, \citenamefont {Bacco}, \citenamefont {Dequal}, \citenamefont
  {Gaiarin}, \citenamefont {Luceri}, \citenamefont {Bianco},\ and\
  \citenamefont {Villoresi}}]{Vallone2015}%
  \BibitemOpen
  \bibfield  {author} {\bibinfo {author} {\bibfnamefont {G.}~\bibnamefont
  {Vallone}}, \bibinfo {author} {\bibfnamefont {D.}~\bibnamefont {Bacco}},
  \bibinfo {author} {\bibfnamefont {D.}~\bibnamefont {Dequal}}, \bibinfo
  {author} {\bibfnamefont {S.}~\bibnamefont {Gaiarin}}, \bibinfo {author}
  {\bibfnamefont {V.}~\bibnamefont {Luceri}}, \bibinfo {author} {\bibfnamefont
  {G.}~\bibnamefont {Bianco}}, \ and\ \bibinfo {author} {\bibfnamefont
  {P.}~\bibnamefont {Villoresi}},\ }\href@noop {} {\bibfield  {journal}
  {\bibinfo  {journal} {Phys. Rev. Lett.}\ }\textbf {\bibinfo {volume} {115}},\
  \bibinfo {pages} {040502} (\bibinfo {year} {2015})}\BibitemShut {NoStop}%
\bibitem [{\citenamefont {Dequal}\ \emph {et~al.}(2016)\citenamefont {Dequal},
  \citenamefont {Vallone}, \citenamefont {Bacco}, \citenamefont {Gaiarin},
  \citenamefont {Luceri}, \citenamefont {Bianco},\ and\ \citenamefont
  {Villoresi}}]{Dequal2016}%
  \BibitemOpen
  \bibfield  {author} {\bibinfo {author} {\bibfnamefont {D.}~\bibnamefont
  {Dequal}}, \bibinfo {author} {\bibfnamefont {G.}~\bibnamefont {Vallone}},
  \bibinfo {author} {\bibfnamefont {D.}~\bibnamefont {Bacco}}, \bibinfo
  {author} {\bibfnamefont {S.}~\bibnamefont {Gaiarin}}, \bibinfo {author}
  {\bibfnamefont {V.}~\bibnamefont {Luceri}}, \bibinfo {author} {\bibfnamefont
  {G.}~\bibnamefont {Bianco}}, \ and\ \bibinfo {author} {\bibfnamefont
  {P.}~\bibnamefont {Villoresi}},\ }\href@noop {} {\bibfield  {journal}
  {\bibinfo  {journal} {Phys. Rev. A}\ }\textbf {\bibinfo {volume} {93}},\
  \bibinfo {pages} {010301(R)} (\bibinfo {year} {2016})}\BibitemShut {NoStop}%
\bibitem [{\citenamefont {Carrasco-Casado}\ \emph {et~al.}(2016)\citenamefont
  {Carrasco-Casado}, \citenamefont {Kunimori}, \citenamefont {Takenaka},
  \citenamefont {Kubo-Oka}, \citenamefont {Akioka}, \citenamefont {Fuse},
  \citenamefont {Koyama}, \citenamefont {Kolev}, \citenamefont {Munemasa},\
  and\ \citenamefont {Toyoshima}}]{Carrasco-Casado2016}%
  \BibitemOpen
  \bibfield  {author} {\bibinfo {author} {\bibfnamefont {A.}~\bibnamefont
  {Carrasco-Casado}}, \bibinfo {author} {\bibfnamefont {H.}~\bibnamefont
  {Kunimori}}, \bibinfo {author} {\bibfnamefont {H.}~\bibnamefont {Takenaka}},
  \bibinfo {author} {\bibfnamefont {T.}~\bibnamefont {Kubo-Oka}}, \bibinfo
  {author} {\bibfnamefont {M.}~\bibnamefont {Akioka}}, \bibinfo {author}
  {\bibfnamefont {T.}~\bibnamefont {Fuse}}, \bibinfo {author} {\bibfnamefont
  {Y.}~\bibnamefont {Koyama}}, \bibinfo {author} {\bibfnamefont
  {D.}~\bibnamefont {Kolev}}, \bibinfo {author} {\bibfnamefont
  {Y.}~\bibnamefont {Munemasa}}, \ and\ \bibinfo {author} {\bibfnamefont
  {M.}~\bibnamefont {Toyoshima}},\ }\href {\doibase 10.1364/OE.24.012254}
  {\bibfield  {journal} {\bibinfo  {journal} {Opt. Express}\ }\textbf {\bibinfo
  {volume} {24}},\ \bibinfo {pages} {12254} (\bibinfo {year}
  {2016})}\BibitemShut {NoStop}%
\bibitem [{\citenamefont {Takenaka}\ \emph {et~al.}(2017)\citenamefont
  {Takenaka}, \citenamefont {Carrasco-Casado}, \citenamefont {Fujiwara},
  \citenamefont {Kitamura}, \citenamefont {Sasaki},\ and\ \citenamefont
  {Toyoshima}}]{Takenaka2017}%
  \BibitemOpen
  \bibfield  {author} {\bibinfo {author} {\bibfnamefont {H.}~\bibnamefont
  {Takenaka}}, \bibinfo {author} {\bibfnamefont {A.}~\bibnamefont
  {Carrasco-Casado}}, \bibinfo {author} {\bibfnamefont {M.}~\bibnamefont
  {Fujiwara}}, \bibinfo {author} {\bibfnamefont {M.}~\bibnamefont {Kitamura}},
  \bibinfo {author} {\bibfnamefont {M.}~\bibnamefont {Sasaki}}, \ and\ \bibinfo
  {author} {\bibfnamefont {M.}~\bibnamefont {Toyoshima}},\ }\href {\doibase
  10.1038/nphoton.2017.107} {\bibfield  {journal} {\bibinfo  {journal} {Nature
  Photonics}\ }\textbf {\bibinfo {volume} {11}},\ \bibinfo {pages} {502}
  (\bibinfo {year} {2017})}\BibitemShut {NoStop}%
\bibitem [{\citenamefont {Agnesi}\ \emph {et~al.}(2018)\citenamefont {Agnesi},
  \citenamefont {Luceri}, \citenamefont {Schiavon}, \citenamefont {Villoresi},
  \citenamefont {Vallone}, \citenamefont {Dequal}, \citenamefont {Santamato},
  \citenamefont {Calderaro}, \citenamefont {Vedovato},\ and\ \citenamefont
  {Bianco}}]{Agnesi2018}%
  \BibitemOpen
  \bibfield  {author} {\bibinfo {author} {\bibfnamefont {C.}~\bibnamefont
  {Agnesi}}, \bibinfo {author} {\bibfnamefont {V.}~\bibnamefont {Luceri}},
  \bibinfo {author} {\bibfnamefont {M.}~\bibnamefont {Schiavon}}, \bibinfo
  {author} {\bibfnamefont {P.}~\bibnamefont {Villoresi}}, \bibinfo {author}
  {\bibfnamefont {G.}~\bibnamefont {Vallone}}, \bibinfo {author} {\bibfnamefont
  {D.}~\bibnamefont {Dequal}}, \bibinfo {author} {\bibfnamefont
  {A.}~\bibnamefont {Santamato}}, \bibinfo {author} {\bibfnamefont
  {L.}~\bibnamefont {Calderaro}}, \bibinfo {author} {\bibfnamefont
  {F.}~\bibnamefont {Vedovato}}, \ and\ \bibinfo {author} {\bibfnamefont
  {G.}~\bibnamefont {Bianco}},\ }\href {\doibase 10.1088/2058-9565/aaefd4}
  {\bibfield  {journal} {\bibinfo  {journal} {Quantum Sci. Technol.}\ }\textbf
  {\bibinfo {volume} {4}},\ \bibinfo {pages} {015012} (\bibinfo {year}
  {2018})}\BibitemShut {NoStop}%
\bibitem [{\citenamefont {Liao}\ \emph {et~al.}(2017)\citenamefont {Liao},
  \citenamefont {Cai}, \citenamefont {Liu}, \citenamefont {Zhang},
  \citenamefont {Li}, \citenamefont {Ren}, \citenamefont {Yin}, \citenamefont
  {Shen}, \citenamefont {Cao}, \citenamefont {Li}, \citenamefont {Li},
  \citenamefont {Chen}, \citenamefont {Sun}, \citenamefont {Jia}, \citenamefont
  {Wu}, \citenamefont {Jiang}, \citenamefont {Wang}, \citenamefont {Huang},
  \citenamefont {Wang}, \citenamefont {Zhou}, \citenamefont {Deng},
  \citenamefont {Xi}, \citenamefont {Ma}, \citenamefont {Hu}, \citenamefont
  {Zhang}, \citenamefont {Chen}, \citenamefont {Liu}, \citenamefont {Wang},
  \citenamefont {Zhu}, \citenamefont {Lu}, \citenamefont {Shu}, \citenamefont
  {Peng}, \citenamefont {Wang},\ and\ \citenamefont {Pan}}]{Liao2017}%
  \BibitemOpen
  \bibfield  {author} {\bibinfo {author} {\bibfnamefont {S.-K.}\ \bibnamefont
  {Liao}}, \bibinfo {author} {\bibfnamefont {W.-Q.}\ \bibnamefont {Cai}},
  \bibinfo {author} {\bibfnamefont {W.-Y.}\ \bibnamefont {Liu}}, \bibinfo
  {author} {\bibfnamefont {L.}~\bibnamefont {Zhang}}, \bibinfo {author}
  {\bibfnamefont {Y.}~\bibnamefont {Li}}, \bibinfo {author} {\bibfnamefont
  {J.-G.}\ \bibnamefont {Ren}}, \bibinfo {author} {\bibfnamefont
  {J.}~\bibnamefont {Yin}}, \bibinfo {author} {\bibfnamefont {Q.}~\bibnamefont
  {Shen}}, \bibinfo {author} {\bibfnamefont {Y.}~\bibnamefont {Cao}}, \bibinfo
  {author} {\bibfnamefont {Z.-P.}\ \bibnamefont {Li}}, \bibinfo {author}
  {\bibfnamefont {F.-Z.}\ \bibnamefont {Li}}, \bibinfo {author} {\bibfnamefont
  {X.-W.}\ \bibnamefont {Chen}}, \bibinfo {author} {\bibfnamefont {L.-H.}\
  \bibnamefont {Sun}}, \bibinfo {author} {\bibfnamefont {J.-J.}\ \bibnamefont
  {Jia}}, \bibinfo {author} {\bibfnamefont {J.-C.}\ \bibnamefont {Wu}},
  \bibinfo {author} {\bibfnamefont {X.-J.}\ \bibnamefont {Jiang}}, \bibinfo
  {author} {\bibfnamefont {J.-F.}\ \bibnamefont {Wang}}, \bibinfo {author}
  {\bibfnamefont {Y.-M.}\ \bibnamefont {Huang}}, \bibinfo {author}
  {\bibfnamefont {Q.}~\bibnamefont {Wang}}, \bibinfo {author} {\bibfnamefont
  {Y.-L.}\ \bibnamefont {Zhou}}, \bibinfo {author} {\bibfnamefont
  {L.}~\bibnamefont {Deng}}, \bibinfo {author} {\bibfnamefont {T.}~\bibnamefont
  {Xi}}, \bibinfo {author} {\bibfnamefont {L.}~\bibnamefont {Ma}}, \bibinfo
  {author} {\bibfnamefont {T.}~\bibnamefont {Hu}}, \bibinfo {author}
  {\bibfnamefont {Q.}~\bibnamefont {Zhang}}, \bibinfo {author} {\bibfnamefont
  {Y.-A.}\ \bibnamefont {Chen}}, \bibinfo {author} {\bibfnamefont {N.-L.}\
  \bibnamefont {Liu}}, \bibinfo {author} {\bibfnamefont {X.-B.}\ \bibnamefont
  {Wang}}, \bibinfo {author} {\bibfnamefont {Z.-C.}\ \bibnamefont {Zhu}},
  \bibinfo {author} {\bibfnamefont {C.-Y.}\ \bibnamefont {Lu}}, \bibinfo
  {author} {\bibfnamefont {R.}~\bibnamefont {Shu}}, \bibinfo {author}
  {\bibfnamefont {C.-Z.}\ \bibnamefont {Peng}}, \bibinfo {author}
  {\bibfnamefont {J.-Y.}\ \bibnamefont {Wang}}, \ and\ \bibinfo {author}
  {\bibfnamefont {J.-W.}\ \bibnamefont {Pan}},\ }\href {\doibase
  10.1038/nature23655} {\bibfield  {journal} {\bibinfo  {journal} {Nature}\
  }\textbf {\bibinfo {volume} {549}},\ \bibinfo {pages} {43} (\bibinfo {year}
  {2017})}\BibitemShut {NoStop}%
\bibitem [{\citenamefont {Yin}\ \emph {et~al.}(2017)\citenamefont {Yin},
  \citenamefont {Cao}, \citenamefont {Li}, \citenamefont {Ren}, \citenamefont
  {Liao}, \citenamefont {Zhang}, \citenamefont {Cai}, \citenamefont {Liu},
  \citenamefont {Li}, \citenamefont {Dai}, \citenamefont {Li}, \citenamefont
  {Huang}, \citenamefont {Deng}, \citenamefont {Li}, \citenamefont {Zhang},
  \citenamefont {Liu}, \citenamefont {Chen}, \citenamefont {Lu}, \citenamefont
  {Shu}, \citenamefont {Peng}, \citenamefont {Wang},\ and\ \citenamefont
  {Pan}}]{Yin2017a}%
  \BibitemOpen
  \bibfield  {author} {\bibinfo {author} {\bibfnamefont {J.}~\bibnamefont
  {Yin}}, \bibinfo {author} {\bibfnamefont {Y.}~\bibnamefont {Cao}}, \bibinfo
  {author} {\bibfnamefont {Y.~H.}\ \bibnamefont {Li}}, \bibinfo {author}
  {\bibfnamefont {J.~G.}\ \bibnamefont {Ren}}, \bibinfo {author} {\bibfnamefont
  {S.~K.}\ \bibnamefont {Liao}}, \bibinfo {author} {\bibfnamefont
  {L.}~\bibnamefont {Zhang}}, \bibinfo {author} {\bibfnamefont {W.~Q.}\
  \bibnamefont {Cai}}, \bibinfo {author} {\bibfnamefont {W.~Y.}\ \bibnamefont
  {Liu}}, \bibinfo {author} {\bibfnamefont {B.}~\bibnamefont {Li}}, \bibinfo
  {author} {\bibfnamefont {H.}~\bibnamefont {Dai}}, \bibinfo {author}
  {\bibfnamefont {M.}~\bibnamefont {Li}}, \bibinfo {author} {\bibfnamefont
  {Y.~M.}\ \bibnamefont {Huang}}, \bibinfo {author} {\bibfnamefont
  {L.}~\bibnamefont {Deng}}, \bibinfo {author} {\bibfnamefont {L.}~\bibnamefont
  {Li}}, \bibinfo {author} {\bibfnamefont {Q.}~\bibnamefont {Zhang}}, \bibinfo
  {author} {\bibfnamefont {N.~L.}\ \bibnamefont {Liu}}, \bibinfo {author}
  {\bibfnamefont {Y.~A.}\ \bibnamefont {Chen}}, \bibinfo {author}
  {\bibfnamefont {C.~Y.}\ \bibnamefont {Lu}}, \bibinfo {author} {\bibfnamefont
  {R.}~\bibnamefont {Shu}}, \bibinfo {author} {\bibfnamefont {C.~Z.}\
  \bibnamefont {Peng}}, \bibinfo {author} {\bibfnamefont {J.~Y.}\ \bibnamefont
  {Wang}}, \ and\ \bibinfo {author} {\bibfnamefont {J.~W.}\ \bibnamefont
  {Pan}},\ }\href {\doibase 10.1103/PhysRevLett.119.200501} {\bibfield
  {journal} {\bibinfo  {journal} {Phys. Rev. Lett.}\ }\textbf {\bibinfo
  {volume} {119}},\ \bibinfo {pages} {200501} (\bibinfo {year}
  {2017})}\BibitemShut {NoStop}%
\bibitem [{\citenamefont {Liao}\ \emph {et~al.}(2018)\citenamefont {Liao},
  \citenamefont {Cai}, \citenamefont {Handsteiner}, \citenamefont {Liu},
  \citenamefont {Yin}, \citenamefont {Zhang}, \citenamefont {Rauch},
  \citenamefont {Fink}, \citenamefont {Ren}, \citenamefont {Liu}, \citenamefont
  {Li}, \citenamefont {Shen}, \citenamefont {Cao}, \citenamefont {Li},
  \citenamefont {Wang}, \citenamefont {Huang}, \citenamefont {Deng},
  \citenamefont {Xi}, \citenamefont {Ma}, \citenamefont {Hu}, \citenamefont
  {Li}, \citenamefont {Liu}, \citenamefont {Koidl}, \citenamefont {Wang},
  \citenamefont {Chen}, \citenamefont {Wang}, \citenamefont {Steindorfer},
  \citenamefont {Kirchner}, \citenamefont {Lu}, \citenamefont {Shu},
  \citenamefont {Ursin}, \citenamefont {Scheidl}, \citenamefont {Peng},
  \citenamefont {Wang}, \citenamefont {Zeilinger},\ and\ \citenamefont
  {Pan}}]{Liao2018}%
  \BibitemOpen
  \bibfield  {author} {\bibinfo {author} {\bibfnamefont {S.~K.}\ \bibnamefont
  {Liao}}, \bibinfo {author} {\bibfnamefont {W.~Q.}\ \bibnamefont {Cai}},
  \bibinfo {author} {\bibfnamefont {J.}~\bibnamefont {Handsteiner}}, \bibinfo
  {author} {\bibfnamefont {B.}~\bibnamefont {Liu}}, \bibinfo {author}
  {\bibfnamefont {J.}~\bibnamefont {Yin}}, \bibinfo {author} {\bibfnamefont
  {L.}~\bibnamefont {Zhang}}, \bibinfo {author} {\bibfnamefont
  {D.}~\bibnamefont {Rauch}}, \bibinfo {author} {\bibfnamefont
  {M.}~\bibnamefont {Fink}}, \bibinfo {author} {\bibfnamefont {J.~G.}\
  \bibnamefont {Ren}}, \bibinfo {author} {\bibfnamefont {W.~Y.}\ \bibnamefont
  {Liu}}, \bibinfo {author} {\bibfnamefont {Y.}~\bibnamefont {Li}}, \bibinfo
  {author} {\bibfnamefont {Q.}~\bibnamefont {Shen}}, \bibinfo {author}
  {\bibfnamefont {Y.}~\bibnamefont {Cao}}, \bibinfo {author} {\bibfnamefont
  {F.~Z.}\ \bibnamefont {Li}}, \bibinfo {author} {\bibfnamefont {J.~F.}\
  \bibnamefont {Wang}}, \bibinfo {author} {\bibfnamefont {Y.~M.}\ \bibnamefont
  {Huang}}, \bibinfo {author} {\bibfnamefont {L.}~\bibnamefont {Deng}},
  \bibinfo {author} {\bibfnamefont {T.}~\bibnamefont {Xi}}, \bibinfo {author}
  {\bibfnamefont {L.}~\bibnamefont {Ma}}, \bibinfo {author} {\bibfnamefont
  {T.}~\bibnamefont {Hu}}, \bibinfo {author} {\bibfnamefont {L.}~\bibnamefont
  {Li}}, \bibinfo {author} {\bibfnamefont {N.~L.}\ \bibnamefont {Liu}},
  \bibinfo {author} {\bibfnamefont {F.}~\bibnamefont {Koidl}}, \bibinfo
  {author} {\bibfnamefont {P.}~\bibnamefont {Wang}}, \bibinfo {author}
  {\bibfnamefont {Y.~A.}\ \bibnamefont {Chen}}, \bibinfo {author}
  {\bibfnamefont {X.~B.}\ \bibnamefont {Wang}}, \bibinfo {author}
  {\bibfnamefont {M.}~\bibnamefont {Steindorfer}}, \bibinfo {author}
  {\bibfnamefont {G.}~\bibnamefont {Kirchner}}, \bibinfo {author}
  {\bibfnamefont {C.~Y.}\ \bibnamefont {Lu}}, \bibinfo {author} {\bibfnamefont
  {R.}~\bibnamefont {Shu}}, \bibinfo {author} {\bibfnamefont {R.}~\bibnamefont
  {Ursin}}, \bibinfo {author} {\bibfnamefont {T.}~\bibnamefont {Scheidl}},
  \bibinfo {author} {\bibfnamefont {C.~Z.}\ \bibnamefont {Peng}}, \bibinfo
  {author} {\bibfnamefont {J.~Y.}\ \bibnamefont {Wang}}, \bibinfo {author}
  {\bibfnamefont {A.}~\bibnamefont {Zeilinger}}, \ and\ \bibinfo {author}
  {\bibfnamefont {J.~W.}\ \bibnamefont {Pan}},\ }\href {\doibase
  10.1103/PhysRevLett.120.030501} {\bibfield  {journal} {\bibinfo  {journal}
  {Phys. Rev. Lett.}\ }\textbf {\bibinfo {volume} {120}},\ \bibinfo {pages}
  {030501} (\bibinfo {year} {2018})}\BibitemShut {NoStop}%
\bibitem [{\citenamefont {Ralph}(2000)}]{Ralph2000}%
  \BibitemOpen
  \bibfield  {author} {\bibinfo {author} {\bibfnamefont {T.~C.}\ \bibnamefont
  {Ralph}},\ }\href {\doibase 10.1103/PhysRevA.61.010303} {\bibfield  {journal}
  {\bibinfo  {journal} {Phys. Rev. A}\ }\textbf {\bibinfo {volume} {61}},\
  \bibinfo {pages} {4} (\bibinfo {year} {2000})}\BibitemShut {NoStop}%
\bibitem [{\citenamefont {Grosshans}\ \emph {et~al.}(2003)\citenamefont
  {Grosshans}, \citenamefont {Van~Assche}, \citenamefont {Wenger},
  \citenamefont {Brouri}, \citenamefont {Cerf},\ and\ \citenamefont
  {Grangier}}]{Grosshans2003}%
  \BibitemOpen
  \bibfield  {author} {\bibinfo {author} {\bibfnamefont {F.}~\bibnamefont
  {Grosshans}}, \bibinfo {author} {\bibfnamefont {G.}~\bibnamefont
  {Van~Assche}}, \bibinfo {author} {\bibfnamefont {J.}~\bibnamefont {Wenger}},
  \bibinfo {author} {\bibfnamefont {R.}~\bibnamefont {Brouri}}, \bibinfo
  {author} {\bibfnamefont {N.~J.}\ \bibnamefont {Cerf}}, \ and\ \bibinfo
  {author} {\bibfnamefont {P.}~\bibnamefont {Grangier}},\ }\href {\doibase
  10.1038/nature01289} {\bibfield  {journal} {\bibinfo  {journal} {Nature}\
  }\textbf {\bibinfo {volume} {421}},\ \bibinfo {pages} {238} (\bibinfo {year}
  {2003})}\BibitemShut {NoStop}%
\bibitem [{\citenamefont {Diamanti}\ and\ \citenamefont
  {Leverrier}(2015)}]{Diamanti2015b}%
  \BibitemOpen
  \bibfield  {author} {\bibinfo {author} {\bibfnamefont {E.}~\bibnamefont
  {Diamanti}}\ and\ \bibinfo {author} {\bibfnamefont {A.}~\bibnamefont
  {Leverrier}},\ }\href {\doibase 10.3390/e17096072} {\bibfield  {journal}
  {\bibinfo  {journal} {Entropy}\ }\textbf {\bibinfo {volume} {17}},\ \bibinfo
  {pages} {6072} (\bibinfo {year} {2015})}\BibitemShut {NoStop}%
\bibitem [{\citenamefont {Laudenbach}\ \emph {et~al.}(2018)\citenamefont
  {Laudenbach}, \citenamefont {Pacher}, \citenamefont {Fung}, \citenamefont
  {Poppe}, \citenamefont {Peev}, \citenamefont {Schrenk}, \citenamefont
  {Hentschel}, \citenamefont {Walther},\ and\ \citenamefont
  {H{\"{u}}bel}}]{Laudenbach2017a}%
  \BibitemOpen
  \bibfield  {author} {\bibinfo {author} {\bibfnamefont {F.}~\bibnamefont
  {Laudenbach}}, \bibinfo {author} {\bibfnamefont {C.}~\bibnamefont {Pacher}},
  \bibinfo {author} {\bibfnamefont {C.-H.~F.}\ \bibnamefont {Fung}}, \bibinfo
  {author} {\bibfnamefont {A.}~\bibnamefont {Poppe}}, \bibinfo {author}
  {\bibfnamefont {M.}~\bibnamefont {Peev}}, \bibinfo {author} {\bibfnamefont
  {B.}~\bibnamefont {Schrenk}}, \bibinfo {author} {\bibfnamefont
  {M.}~\bibnamefont {Hentschel}}, \bibinfo {author} {\bibfnamefont
  {P.}~\bibnamefont {Walther}}, \ and\ \bibinfo {author} {\bibfnamefont
  {H.}~\bibnamefont {H{\"{u}}bel}},\ }\href {\doibase 10.1002/qute.201800011}
  {\bibfield  {journal} {\bibinfo  {journal} {Adv. Quantum Technol.}\ }\textbf
  {\bibinfo {volume} {1}},\ \bibinfo {pages} {1800011} (\bibinfo {year}
  {2018})}\BibitemShut {NoStop}%
\bibitem [{\citenamefont {Vasylyev}\ \emph {et~al.}(2012)\citenamefont
  {Vasylyev}, \citenamefont {Semenov},\ and\ \citenamefont
  {Vogel}}]{Vasylyev2012}%
  \BibitemOpen
  \bibfield  {author} {\bibinfo {author} {\bibfnamefont {D.~Y.}\ \bibnamefont
  {Vasylyev}}, \bibinfo {author} {\bibfnamefont {A.~A.}\ \bibnamefont
  {Semenov}}, \ and\ \bibinfo {author} {\bibfnamefont {W.}~\bibnamefont
  {Vogel}},\ }\href {\doibase 10.1103/PhysRevLett.108.220501} {\bibfield
  {journal} {\bibinfo  {journal} {Phys. Rev. Lett.}\ }\textbf {\bibinfo
  {volume} {108}},\ \bibinfo {pages} {220501} (\bibinfo {year}
  {2012})}\BibitemShut {NoStop}%
\bibitem [{\citenamefont {Semenov}\ \emph {et~al.}(2012)\citenamefont
  {Semenov}, \citenamefont {T{\"{o}}ppel}, \citenamefont {Vasylyev},
  \citenamefont {Gomonay},\ and\ \citenamefont {Vogel}}]{Semenov2012}%
  \BibitemOpen
  \bibfield  {author} {\bibinfo {author} {\bibfnamefont {A.~A.}\ \bibnamefont
  {Semenov}}, \bibinfo {author} {\bibfnamefont {F.}~\bibnamefont
  {T{\"{o}}ppel}}, \bibinfo {author} {\bibfnamefont {D.~Y.}\ \bibnamefont
  {Vasylyev}}, \bibinfo {author} {\bibfnamefont {H.~V.}\ \bibnamefont
  {Gomonay}}, \ and\ \bibinfo {author} {\bibfnamefont {W.}~\bibnamefont
  {Vogel}},\ }\href {\doibase 10.1103/PhysRevA.85.013826} {\bibfield  {journal}
  {\bibinfo  {journal} {Phys. Rev. A}\ }\textbf {\bibinfo {volume} {85}},\
  \bibinfo {pages} {013826} (\bibinfo {year} {2012})}\BibitemShut {NoStop}%
\bibitem [{\citenamefont {Wang}\ \emph {et~al.}(2018)\citenamefont {Wang},
  \citenamefont {Huang}, \citenamefont {Wang},\ and\ \citenamefont
  {Zeng}}]{Wang2018}%
  \BibitemOpen
  \bibfield  {author} {\bibinfo {author} {\bibfnamefont {S.}~\bibnamefont
  {Wang}}, \bibinfo {author} {\bibfnamefont {P.}~\bibnamefont {Huang}},
  \bibinfo {author} {\bibfnamefont {T.}~\bibnamefont {Wang}}, \ and\ \bibinfo
  {author} {\bibfnamefont {G.}~\bibnamefont {Zeng}},\ }\href {\doibase
  10.1016/j.polymer.2005.07.060} {\bibfield  {journal} {\bibinfo  {journal}
  {New J. Phys.}\ }\textbf {\bibinfo {volume} {20}},\ \bibinfo {pages} {083037}
  (\bibinfo {year} {2018})}\BibitemShut {NoStop}%
\bibitem [{\citenamefont {Ruppert}\ \emph {et~al.}(2019)\citenamefont
  {Ruppert}, \citenamefont {Peuntinger}, \citenamefont {Heim}, \citenamefont
  {G\"{u}nthner}, \citenamefont {Usenko}, \citenamefont {Elser}, \citenamefont
  {Leuchs}, \citenamefont {Filip},\ and\ \citenamefont
  {Marquardt}}]{Ruppert2019}%
  \BibitemOpen
  \bibfield  {author} {\bibinfo {author} {\bibfnamefont {L.}~\bibnamefont
  {Ruppert}}, \bibinfo {author} {\bibfnamefont {C.}~\bibnamefont {Peuntinger}},
  \bibinfo {author} {\bibfnamefont {B.}~\bibnamefont {Heim}}, \bibinfo {author}
  {\bibfnamefont {K.}~\bibnamefont {G\"{u}nthner}}, \bibinfo {author}
  {\bibfnamefont {V.~C.}\ \bibnamefont {Usenko}}, \bibinfo {author}
  {\bibfnamefont {D.}~\bibnamefont {Elser}}, \bibinfo {author} {\bibfnamefont
  {G.}~\bibnamefont {Leuchs}}, \bibinfo {author} {\bibfnamefont
  {R.}~\bibnamefont {Filip}}, \ and\ \bibinfo {author} {\bibfnamefont
  {C.}~\bibnamefont {Marquardt}},\ }\href {\doibase 10.1088/1367-2630/ab5dd3}
  {\bibfield  {journal} {\bibinfo  {journal} {New J. Phys.}\ }\textbf {\bibinfo
  {volume} {21}},\ \bibinfo {pages} {123036} (\bibinfo {year}
  {2019})}\BibitemShut {NoStop}%
\bibitem [{\citenamefont {Heim}\ \emph {et~al.}(2014)\citenamefont {Heim},
  \citenamefont {Peuntinger}, \citenamefont {Killoran}, \citenamefont {Khan},
  \citenamefont {Wittmann}, \citenamefont {Marquardt},\ and\ \citenamefont
  {Leuchs}}]{Heim2014}%
  \BibitemOpen
  \bibfield  {author} {\bibinfo {author} {\bibfnamefont {B.}~\bibnamefont
  {Heim}}, \bibinfo {author} {\bibfnamefont {C.}~\bibnamefont {Peuntinger}},
  \bibinfo {author} {\bibfnamefont {N.}~\bibnamefont {Killoran}}, \bibinfo
  {author} {\bibfnamefont {I.}~\bibnamefont {Khan}}, \bibinfo {author}
  {\bibfnamefont {C.}~\bibnamefont {Wittmann}}, \bibinfo {author}
  {\bibfnamefont {C.}~\bibnamefont {Marquardt}}, \ and\ \bibinfo {author}
  {\bibfnamefont {G.}~\bibnamefont {Leuchs}},\ }\href {\doibase
  10.1088/1367-2630/16/11/113018} {\bibfield  {journal} {\bibinfo  {journal}
  {New J. Phys.}\ }\textbf {\bibinfo {volume} {16}},\ \bibinfo {pages} {113018}
  (\bibinfo {year} {2014})}\BibitemShut {NoStop}%
\bibitem [{\citenamefont {G{\"{u}}nthner}\ \emph {et~al.}(2016)\citenamefont
  {G{\"{u}}nthner}, \citenamefont {Khan}, \citenamefont {Elser}, \citenamefont
  {Stiller}, \citenamefont {Bayraktar}, \citenamefont {M{\"{u}}ller},
  \citenamefont {Saucke}, \citenamefont {Tr{\"{o}}ndle}, \citenamefont {Heine},
  \citenamefont {Seel}, \citenamefont {Greulich}, \citenamefont {Zech},
  \citenamefont {G{\"{u}}tlich}, \citenamefont {Philipp-May}, \citenamefont
  {Marquardt},\ and\ \citenamefont {Leuchs}}]{Gunter2017}%
  \BibitemOpen
  \bibfield  {author} {\bibinfo {author} {\bibfnamefont {K.}~\bibnamefont
  {G{\"{u}}nthner}}, \bibinfo {author} {\bibfnamefont {I.}~\bibnamefont
  {Khan}}, \bibinfo {author} {\bibfnamefont {D.}~\bibnamefont {Elser}},
  \bibinfo {author} {\bibfnamefont {B.}~\bibnamefont {Stiller}}, \bibinfo
  {author} {\bibfnamefont {Ã.}~\bibnamefont {Bayraktar}}, \bibinfo {author}
  {\bibfnamefont {C.~R.}\ \bibnamefont {M{\"{u}}ller}}, \bibinfo {author}
  {\bibfnamefont {K.}~\bibnamefont {Saucke}}, \bibinfo {author} {\bibfnamefont
  {D.}~\bibnamefont {Tr{\"{o}}ndle}}, \bibinfo {author} {\bibfnamefont
  {F.}~\bibnamefont {Heine}}, \bibinfo {author} {\bibfnamefont
  {S.}~\bibnamefont {Seel}}, \bibinfo {author} {\bibfnamefont {P.}~\bibnamefont
  {Greulich}}, \bibinfo {author} {\bibfnamefont {H.}~\bibnamefont {Zech}},
  \bibinfo {author} {\bibfnamefont {B.}~\bibnamefont {G{\"{u}}tlich}}, \bibinfo
  {author} {\bibfnamefont {S.}~\bibnamefont {Philipp-May}}, \bibinfo {author}
  {\bibfnamefont {C.}~\bibnamefont {Marquardt}}, \ and\ \bibinfo {author}
  {\bibfnamefont {G.}~\bibnamefont {Leuchs}},\ }\href {\doibase
  10.1364/OPTICA.4.000611} {\bibfield  {journal} {\bibinfo  {journal} {Optica}\
  }\textbf {\bibinfo {volume} {4}},\ \bibinfo {pages} {611} (\bibinfo {year}
  {2016})}\BibitemShut {NoStop}%
\bibitem [{\citenamefont {Hosseinidehaj}\ \emph {et~al.}(2019)\citenamefont
  {Hosseinidehaj}, \citenamefont {Babar}, \citenamefont {Malaney},
  \citenamefont {Ng},\ and\ \citenamefont {Hanzo}}]{Hosseinidehaj2019}%
  \BibitemOpen
  \bibfield  {author} {\bibinfo {author} {\bibfnamefont {N.}~\bibnamefont
  {Hosseinidehaj}}, \bibinfo {author} {\bibfnamefont {Z.}~\bibnamefont
  {Babar}}, \bibinfo {author} {\bibfnamefont {R.}~\bibnamefont {Malaney}},
  \bibinfo {author} {\bibfnamefont {S.~X.}\ \bibnamefont {Ng}}, \ and\ \bibinfo
  {author} {\bibfnamefont {L.}~\bibnamefont {Hanzo}},\ }\href {\doibase
  10.1109/comst.2018.2864557} {\bibfield  {journal} {\bibinfo  {journal}
  {{IEEE} Communications Surveys {\&} Tutorials}\ }\textbf {\bibinfo {volume}
  {21}},\ \bibinfo {pages} {881} (\bibinfo {year} {2019})}\BibitemShut
  {NoStop}%
\bibitem [{\citenamefont {Guo}\ \emph {et~al.}(2018)\citenamefont {Guo},
  \citenamefont {Xie}, \citenamefont {Huang}, \citenamefont {Li}, \citenamefont
  {Zhang}, \citenamefont {Huang},\ and\ \citenamefont {Zeng}}]{Guo2018}%
  \BibitemOpen
  \bibfield  {author} {\bibinfo {author} {\bibfnamefont {Y.}~\bibnamefont
  {Guo}}, \bibinfo {author} {\bibfnamefont {C.}~\bibnamefont {Xie}}, \bibinfo
  {author} {\bibfnamefont {P.}~\bibnamefont {Huang}}, \bibinfo {author}
  {\bibfnamefont {J.}~\bibnamefont {Li}}, \bibinfo {author} {\bibfnamefont
  {L.}~\bibnamefont {Zhang}}, \bibinfo {author} {\bibfnamefont
  {D.}~\bibnamefont {Huang}}, \ and\ \bibinfo {author} {\bibfnamefont
  {G.}~\bibnamefont {Zeng}},\ }\href {\doibase 10.1103/physreva.97.052326}
  {\bibfield  {journal} {\bibinfo  {journal} {Phys. Rev. A}\ }\textbf {\bibinfo
  {volume} {97}} (\bibinfo {year} {2018}),\
  10.1103/physreva.97.052326}\BibitemShut {NoStop}%
\bibitem [{\citenamefont {Villasenor}\ \emph {et~al.}(2020)\citenamefont
  {Villasenor}, \citenamefont {Malaney}, \citenamefont {Mudge},\ and\
  \citenamefont {Grant}}]{villasenor2020atmospheric}%
  \BibitemOpen
  \bibfield  {author} {\bibinfo {author} {\bibfnamefont {E.}~\bibnamefont
  {Villasenor}}, \bibinfo {author} {\bibfnamefont {R.}~\bibnamefont {Malaney}},
  \bibinfo {author} {\bibfnamefont {K.~A.}\ \bibnamefont {Mudge}}, \ and\
  \bibinfo {author} {\bibfnamefont {K.~J.}\ \bibnamefont {Grant}},\ }\href@noop
  {} {\enquote {\bibinfo {title} {Atmospheric effects on satellite-to-ground
  quantum key distribution using coherent states},}\ } (\bibinfo {year}
  {2020}),\ \Eprint {http://arxiv.org/abs/2005.10465} {arXiv:2005.10465
  [quant-ph]} \BibitemShut {NoStop}%
\bibitem [{\citenamefont {Kish}\ \emph {et~al.}(2020)\citenamefont {Kish},
  \citenamefont {Villasenor}, \citenamefont {Malaney}, \citenamefont {Mudge},\
  and\ \citenamefont {Grant}}]{kish2020feasibility}%
  \BibitemOpen
  \bibfield  {author} {\bibinfo {author} {\bibfnamefont {S.}~\bibnamefont
  {Kish}}, \bibinfo {author} {\bibfnamefont {E.}~\bibnamefont {Villasenor}},
  \bibinfo {author} {\bibfnamefont {R.}~\bibnamefont {Malaney}}, \bibinfo
  {author} {\bibfnamefont {K.}~\bibnamefont {Mudge}}, \ and\ \bibinfo {author}
  {\bibfnamefont {K.}~\bibnamefont {Grant}},\ }\href@noop {} {\enquote
  {\bibinfo {title} {Feasibility assessment for practical continuous variable
  quantum key distribution over the satellite-to-earth channel},}\ } (\bibinfo
  {year} {2020}),\ \Eprint {http://arxiv.org/abs/2005.10468} {arXiv:2005.10468
  [quant-ph]} \BibitemShut {NoStop}%
\bibitem [{\citenamefont {Yin}\ \emph {et~al.}(2020)\citenamefont {Yin},
  \citenamefont {Li}, \citenamefont {Liao}, \citenamefont {Yang}, \citenamefont
  {Cao}, \citenamefont {Zhang}, \citenamefont {Ren}, \citenamefont {Cai},
  \citenamefont {Liu}, \citenamefont {Li}, \citenamefont {Shu}, \citenamefont
  {Huang}, \citenamefont {Deng}, \citenamefont {Li}, \citenamefont {Zhang},
  \citenamefont {Liu}, \citenamefont {Chen}, \citenamefont {Lu}, \citenamefont
  {Wang}, \citenamefont {Xu}, \citenamefont {Wang}, \citenamefont {Peng},
  \citenamefont {Ekert},\ and\ \citenamefont {Pan}}]{Yin2020}%
  \BibitemOpen
  \bibfield  {author} {\bibinfo {author} {\bibfnamefont {J.}~\bibnamefont
  {Yin}}, \bibinfo {author} {\bibfnamefont {Y.-H.}\ \bibnamefont {Li}},
  \bibinfo {author} {\bibfnamefont {S.-K.}\ \bibnamefont {Liao}}, \bibinfo
  {author} {\bibfnamefont {M.}~\bibnamefont {Yang}}, \bibinfo {author}
  {\bibfnamefont {Y.}~\bibnamefont {Cao}}, \bibinfo {author} {\bibfnamefont
  {L.}~\bibnamefont {Zhang}}, \bibinfo {author} {\bibfnamefont {J.-G.}\
  \bibnamefont {Ren}}, \bibinfo {author} {\bibfnamefont {W.-Q.}\ \bibnamefont
  {Cai}}, \bibinfo {author} {\bibfnamefont {W.-Y.}\ \bibnamefont {Liu}},
  \bibinfo {author} {\bibfnamefont {S.-L.}\ \bibnamefont {Li}}, \bibinfo
  {author} {\bibfnamefont {R.}~\bibnamefont {Shu}}, \bibinfo {author}
  {\bibfnamefont {Y.-M.}\ \bibnamefont {Huang}}, \bibinfo {author}
  {\bibfnamefont {L.}~\bibnamefont {Deng}}, \bibinfo {author} {\bibfnamefont
  {L.}~\bibnamefont {Li}}, \bibinfo {author} {\bibfnamefont {Q.}~\bibnamefont
  {Zhang}}, \bibinfo {author} {\bibfnamefont {N.-L.}\ \bibnamefont {Liu}},
  \bibinfo {author} {\bibfnamefont {Y.-A.}\ \bibnamefont {Chen}}, \bibinfo
  {author} {\bibfnamefont {C.-Y.}\ \bibnamefont {Lu}}, \bibinfo {author}
  {\bibfnamefont {X.-B.}\ \bibnamefont {Wang}}, \bibinfo {author}
  {\bibfnamefont {F.}~\bibnamefont {Xu}}, \bibinfo {author} {\bibfnamefont
  {J.-Y.}\ \bibnamefont {Wang}}, \bibinfo {author} {\bibfnamefont {C.-Z.}\
  \bibnamefont {Peng}}, \bibinfo {author} {\bibfnamefont {A.~K.}\ \bibnamefont
  {Ekert}}, \ and\ \bibinfo {author} {\bibfnamefont {J.-W.}\ \bibnamefont
  {Pan}},\ }\href {\doibase 10.1038/s41586-020-2401-y} {\bibfield  {journal}
  {\bibinfo  {journal} {Nature}\ }\textbf {\bibinfo {volume} {582}},\ \bibinfo
  {pages} {501} (\bibinfo {year} {2020})}\BibitemShut {NoStop}%
\bibitem [{\citenamefont {Bourgoin}\ \emph {et~al.}(2013)\citenamefont
  {Bourgoin}, \citenamefont {Meyer-Scott}, \citenamefont {Higgins},
  \citenamefont {Helou}, \citenamefont {Erven}, \citenamefont {H\"ubel},
  \citenamefont {Kumar}, \citenamefont {Hudson}, \citenamefont {D'Souza},
  \citenamefont {Girard}, \citenamefont {Laflamme},\ and\ \citenamefont
  {Jennewein}}]{Bourgoin2013}%
  \BibitemOpen
  \bibfield  {author} {\bibinfo {author} {\bibfnamefont {J.-P.}\ \bibnamefont
  {Bourgoin}}, \bibinfo {author} {\bibfnamefont {E.}~\bibnamefont
  {Meyer-Scott}}, \bibinfo {author} {\bibfnamefont {B.~L.}\ \bibnamefont
  {Higgins}}, \bibinfo {author} {\bibfnamefont {B.}~\bibnamefont {Helou}},
  \bibinfo {author} {\bibfnamefont {C.}~\bibnamefont {Erven}}, \bibinfo
  {author} {\bibfnamefont {H.}~\bibnamefont {H\"ubel}}, \bibinfo {author}
  {\bibfnamefont {B.}~\bibnamefont {Kumar}}, \bibinfo {author} {\bibfnamefont
  {D.}~\bibnamefont {Hudson}}, \bibinfo {author} {\bibfnamefont
  {I.}~\bibnamefont {D'Souza}}, \bibinfo {author} {\bibfnamefont
  {R.}~\bibnamefont {Girard}}, \bibinfo {author} {\bibfnamefont
  {R.}~\bibnamefont {Laflamme}}, \ and\ \bibinfo {author} {\bibfnamefont
  {T.}~\bibnamefont {Jennewein}},\ }\href@noop {} {\bibfield  {journal}
  {\bibinfo  {journal} {New J. Phys.}\ }\textbf {\bibinfo {volume} {15}},\
  \bibinfo {pages} {023006} (\bibinfo {year} {2013})}\BibitemShut {NoStop}%
\bibitem [{\citenamefont {Vasylyev}\ \emph {et~al.}(2019)\citenamefont
  {Vasylyev}, \citenamefont {Vogel},\ and\ \citenamefont
  {Moll}}]{Vasylyev2019}%
  \BibitemOpen
  \bibfield  {author} {\bibinfo {author} {\bibfnamefont {D.}~\bibnamefont
  {Vasylyev}}, \bibinfo {author} {\bibfnamefont {W.}~\bibnamefont {Vogel}}, \
  and\ \bibinfo {author} {\bibfnamefont {F.}~\bibnamefont {Moll}},\ }\href@noop
  {} {\bibfield  {journal} {\bibinfo  {journal} {Phys. Rev. A}\ }\textbf
  {\bibinfo {volume} {99}},\ \bibinfo {pages} {053830} (\bibinfo {year}
  {2019})}\BibitemShut {NoStop}%
\bibitem [{\citenamefont {Tomasi}\ and\ \citenamefont
  {Petkov}(2014)}]{Tomasi2014}%
  \BibitemOpen
  \bibfield  {author} {\bibinfo {author} {\bibfnamefont {C.}~\bibnamefont
  {Tomasi}}\ and\ \bibinfo {author} {\bibfnamefont {B.~H.}\ \bibnamefont
  {Petkov}},\ }\href {\doibase 10.1002/2013jd020600} {\bibfield  {journal}
  {\bibinfo  {journal} {Journal of Geophysical Research: Atmospheres}\ }\textbf
  {\bibinfo {volume} {119}},\ \bibinfo {pages} {1363} (\bibinfo {year}
  {2014})}\BibitemShut {NoStop}%
\bibitem [{\citenamefont {Berk}\ \emph {et~al.}(2014)\citenamefont {Berk},
  \citenamefont {Conforti}, \citenamefont {Kennett}, \citenamefont {Perkins},
  \citenamefont {Hawes},\ and\ \citenamefont {van~den Bosch}}]{Berk2014}%
  \BibitemOpen
  \bibfield  {author} {\bibinfo {author} {\bibfnamefont {A.}~\bibnamefont
  {Berk}}, \bibinfo {author} {\bibfnamefont {P.}~\bibnamefont {Conforti}},
  \bibinfo {author} {\bibfnamefont {R.}~\bibnamefont {Kennett}}, \bibinfo
  {author} {\bibfnamefont {T.}~\bibnamefont {Perkins}}, \bibinfo {author}
  {\bibfnamefont {F.}~\bibnamefont {Hawes}}, \ and\ \bibinfo {author}
  {\bibfnamefont {J.}~\bibnamefont {van~den Bosch}},\ }in\ \href {\doibase
  10.1117/12.2050433} {\emph {\bibinfo {booktitle} {Algorithms and Technologies
  for Multispectral, Hyperspectral, and Ultraspectral Imagery {XX}}}},\
  \bibinfo {editor} {edited by\ \bibinfo {editor} {\bibfnamefont
  {M.}~\bibnamefont {Velez-Reyes}}\ and\ \bibinfo {editor} {\bibfnamefont
  {F.~A.}\ \bibnamefont {Kruse}}}\ (\bibinfo  {publisher} {{SPIE}},\ \bibinfo
  {year} {2014})\BibitemShut {NoStop}%
\bibitem [{\citenamefont {Inc.}(2020)}]{Modtran}%
  \BibitemOpen
  \bibfield  {author} {\bibinfo {author} {\bibfnamefont {S.~S.}\ \bibnamefont
  {Inc.}},\ }\href {http://modtran.spectral.com/modtran_home} {\enquote
  {\bibinfo {title} {Modtran web app},}\ } (\bibinfo {year} {2020})\BibitemShut
  {NoStop}%
\bibitem [{\citenamefont {Tyson}(2011)}]{Tyson2011}%
  \BibitemOpen
  \bibfield  {author} {\bibinfo {author} {\bibfnamefont {R.}~\bibnamefont
  {Tyson}},\ }\href@noop {} {\emph {\bibinfo {title} {Principles of Adaptive
  Optics}}}\ (\bibinfo  {publisher} {CRC Press, Boca Raton, 3rd edition},\
  \bibinfo {year} {2011})\BibitemShut {NoStop}%
\bibitem [{\citenamefont {Jovanovic}\ \emph {et~al.}(2017)\citenamefont
  {Jovanovic}, \citenamefont {Schwab}, \citenamefont {Guyon}, \citenamefont
  {Lozi}, \citenamefont {Cvetojevic}, \citenamefont {Martinache}, \citenamefont
  {Leon-Saval}, \citenamefont {Norris}, \citenamefont {Gross}, \citenamefont
  {Doughty}, \citenamefont {Currie},\ and\ \citenamefont
  {Takato}}]{Jovanovic2017a}%
  \BibitemOpen
  \bibfield  {author} {\bibinfo {author} {\bibfnamefont {N.}~\bibnamefont
  {Jovanovic}}, \bibinfo {author} {\bibfnamefont {C.}~\bibnamefont {Schwab}},
  \bibinfo {author} {\bibfnamefont {O.}~\bibnamefont {Guyon}}, \bibinfo
  {author} {\bibfnamefont {J.}~\bibnamefont {Lozi}}, \bibinfo {author}
  {\bibfnamefont {N.}~\bibnamefont {Cvetojevic}}, \bibinfo {author}
  {\bibfnamefont {F.}~\bibnamefont {Martinache}}, \bibinfo {author}
  {\bibfnamefont {S.}~\bibnamefont {Leon-Saval}}, \bibinfo {author}
  {\bibfnamefont {B.}~\bibnamefont {Norris}}, \bibinfo {author} {\bibfnamefont
  {S.}~\bibnamefont {Gross}}, \bibinfo {author} {\bibfnamefont
  {D.}~\bibnamefont {Doughty}}, \bibinfo {author} {\bibfnamefont
  {T.}~\bibnamefont {Currie}}, \ and\ \bibinfo {author} {\bibfnamefont
  {N.}~\bibnamefont {Takato}},\ }\href {\doibase 10.1051/0004-6361/201630351}
  {\bibfield  {journal} {\bibinfo  {journal} {Astronomy \& Astrophysics}\
  }\textbf {\bibinfo {volume} {604}},\ \bibinfo {pages} {A122} (\bibinfo {year}
  {2017})}\BibitemShut {NoStop}%
\bibitem [{\citenamefont {Fante}(1980)}]{Fante1980}%
  \BibitemOpen
  \bibfield  {author} {\bibinfo {author} {\bibfnamefont {R.}~\bibnamefont
  {Fante}},\ }\href {\doibase 10.1109/proc.1980.11882} {\bibfield  {journal}
  {\bibinfo  {journal} {Proceedings of the {IEEE}}\ }\textbf {\bibinfo {volume}
  {68}},\ \bibinfo {pages} {1424} (\bibinfo {year} {1980})}\BibitemShut
  {NoStop}%
\bibitem [{\citenamefont {{Liorni}}\ \emph {et~al.}(2019)\citenamefont
  {{Liorni}}, \citenamefont {{Kampermann}},\ and\ \citenamefont
  {{Bruss}}}]{Liorni2019}%
  \BibitemOpen
  \bibfield  {author} {\bibinfo {author} {\bibfnamefont {C.}~\bibnamefont
  {{Liorni}}}, \bibinfo {author} {\bibfnamefont {H.}~\bibnamefont
  {{Kampermann}}}, \ and\ \bibinfo {author} {\bibfnamefont {D.}~\bibnamefont
  {{Bruss}}},\ }\href@noop {} {\bibfield  {journal} {\bibinfo  {journal} {New
  J. Phys.}\ }\textbf {\bibinfo {volume} {21}},\ \bibinfo {pages} {093055}
  (\bibinfo {year} {2019})}\BibitemShut {NoStop}%
\bibitem [{\citenamefont {Weedbrook}\ \emph {et~al.}(2004)\citenamefont
  {Weedbrook}, \citenamefont {Lance}, \citenamefont {Bowen}, \citenamefont
  {Symul}, \citenamefont {Ralph},\ and\ \citenamefont {Lam}}]{Weedbrook2004}%
  \BibitemOpen
  \bibfield  {author} {\bibinfo {author} {\bibfnamefont {C.}~\bibnamefont
  {Weedbrook}}, \bibinfo {author} {\bibfnamefont {A.~M.}\ \bibnamefont
  {Lance}}, \bibinfo {author} {\bibfnamefont {W.~P.}\ \bibnamefont {Bowen}},
  \bibinfo {author} {\bibfnamefont {T.}~\bibnamefont {Symul}}, \bibinfo
  {author} {\bibfnamefont {T.~C.}\ \bibnamefont {Ralph}}, \ and\ \bibinfo
  {author} {\bibfnamefont {P.~K.}\ \bibnamefont {Lam}},\ }\href {\doibase
  10.1103/physrevlett.93.170504} {\bibfield  {journal} {\bibinfo  {journal}
  {Phys. Rev. Lett.}\ }\textbf {\bibinfo {volume} {93}},\ \bibinfo {pages}
  {170504} (\bibinfo {year} {2004})}\BibitemShut {NoStop}%
\bibitem [{Note1()}]{Note1}%
  \BibitemOpen
  \bibinfo {note} {In practice, Bob splits the signal onto a balanced
  beamsplitter then measures the $\protect \hat {q} = b+b^\dagger $ quadrature
  of one output mode and the $\protect \hat {p} = i(b^\dagger - b)$ quadrature
  of the second output mode. He then stores the first measurement outcome in
  the variable $Y_{2k-1}$ and the second outcome in $Y_{2k}$.}\BibitemShut
  {Stop}%
\bibitem [{\citenamefont {Grosshans}\ and\ \citenamefont
  {Grangier}(2002)}]{Grosshans2002Arxiv}%
  \BibitemOpen
  \bibfield  {author} {\bibinfo {author} {\bibfnamefont {F.}~\bibnamefont
  {Grosshans}}\ and\ \bibinfo {author} {\bibfnamefont {P.}~\bibnamefont
  {Grangier}},\ }\href@noop {} {\enquote {\bibinfo {title} {Reverse
  reconciliation protocols for quantum cryptography with continuous
  variables},}\ } (\bibinfo {year} {2002}),\ \Eprint
  {http://arxiv.org/abs/0204127} {arXiv:0204127 [quant-ph]} \BibitemShut
  {NoStop}%
\bibitem [{\citenamefont {Devetak}\ and\ \citenamefont
  {Winter}(2005)}]{Devetak2005}%
  \BibitemOpen
  \bibfield  {author} {\bibinfo {author} {\bibfnamefont {I.}~\bibnamefont
  {Devetak}}\ and\ \bibinfo {author} {\bibfnamefont {A.}~\bibnamefont
  {Winter}},\ }\href {\doibase 10.1098/rspa.2004.1372} {\bibfield  {journal}
  {\bibinfo  {journal} {Proceedings of the Royal Society A: Mathematical,
  Physical and Engineering Sciences}\ }\textbf {\bibinfo {volume} {461}},\
  \bibinfo {pages} {207} (\bibinfo {year} {2005})}\BibitemShut {NoStop}%
\bibitem [{\citenamefont {Renner}\ and\ \citenamefont
  {Cirac}(2009)}]{Renner2009}%
  \BibitemOpen
  \bibfield  {author} {\bibinfo {author} {\bibfnamefont {R.}~\bibnamefont
  {Renner}}\ and\ \bibinfo {author} {\bibfnamefont {J.~I.}\ \bibnamefont
  {Cirac}},\ }\href {\doibase 10.1103/PhysRevLett.102.110504} {\bibfield
  {journal} {\bibinfo  {journal} {Phys. Rev. Lett.}\ }\textbf {\bibinfo
  {volume} {102}},\ \bibinfo {pages} {110504} (\bibinfo {year}
  {2009})}\BibitemShut {NoStop}%
\bibitem [{\citenamefont {Christandl}\ \emph {et~al.}(2009)\citenamefont
  {Christandl}, \citenamefont {Konig},\ and\ \citenamefont
  {Renner}}]{Christandl2009}%
  \BibitemOpen
  \bibfield  {author} {\bibinfo {author} {\bibfnamefont {M.}~\bibnamefont
  {Christandl}}, \bibinfo {author} {\bibfnamefont {R.}~\bibnamefont {Konig}}, \
  and\ \bibinfo {author} {\bibfnamefont {R.}~\bibnamefont {Renner}},\ }\href
  {\doibase 10.1103/physrevlett.102.020504} {\bibfield  {journal} {\bibinfo
  {journal} {Phys. Rev. Lett.}\ }\textbf {\bibinfo {volume} {102}},\ \bibinfo
  {pages} {020504} (\bibinfo {year} {2009})}\BibitemShut {NoStop}%
\bibitem [{\citenamefont {Leverrier}(2017)}]{Leverrier2017}%
  \BibitemOpen
  \bibfield  {author} {\bibinfo {author} {\bibfnamefont {A.}~\bibnamefont
  {Leverrier}},\ }\href {\doibase 10.1103/PhysRevLett.118.200501} {\bibfield
  {journal} {\bibinfo  {journal} {Phys. Rev. Lett.}\ }\textbf {\bibinfo
  {volume} {118}},\ \bibinfo {pages} {200501} (\bibinfo {year}
  {2017})}\BibitemShut {NoStop}%
\bibitem [{\citenamefont {Usenko}\ \emph {et~al.}(2012)\citenamefont {Usenko},
  \citenamefont {Heim}, \citenamefont {Peuntinger}, \citenamefont {Wittmann},
  \citenamefont {Marquardt}, \citenamefont {Leuchs},\ and\ \citenamefont
  {Filip}}]{Usenko2012}%
  \BibitemOpen
  \bibfield  {author} {\bibinfo {author} {\bibfnamefont {V.~C.}\ \bibnamefont
  {Usenko}}, \bibinfo {author} {\bibfnamefont {B.}~\bibnamefont {Heim}},
  \bibinfo {author} {\bibfnamefont {C.}~\bibnamefont {Peuntinger}}, \bibinfo
  {author} {\bibfnamefont {C.}~\bibnamefont {Wittmann}}, \bibinfo {author}
  {\bibfnamefont {C.}~\bibnamefont {Marquardt}}, \bibinfo {author}
  {\bibfnamefont {G.}~\bibnamefont {Leuchs}}, \ and\ \bibinfo {author}
  {\bibfnamefont {R.}~\bibnamefont {Filip}},\ }\href {\doibase
  10.1088/1367-2630/14/9/093048} {\bibfield  {journal} {\bibinfo  {journal}
  {New J. Phys.}\ }\textbf {\bibinfo {volume} {14}},\ \bibinfo {pages} {093048}
  (\bibinfo {year} {2012})}\BibitemShut {NoStop}%
\bibitem [{\citenamefont {Papanastasiou}\ \emph {et~al.}(2018)\citenamefont
  {Papanastasiou}, \citenamefont {Weedbrook},\ and\ \citenamefont
  {Pirandola}}]{Papanastasiou2018}%
  \BibitemOpen
  \bibfield  {author} {\bibinfo {author} {\bibfnamefont {P.}~\bibnamefont
  {Papanastasiou}}, \bibinfo {author} {\bibfnamefont {C.}~\bibnamefont
  {Weedbrook}}, \ and\ \bibinfo {author} {\bibfnamefont {S.}~\bibnamefont
  {Pirandola}},\ }\href {\doibase 10.1103/PhysRevA.97.032311} {\bibfield
  {journal} {\bibinfo  {journal} {Phys. Rev. A}\ }\textbf {\bibinfo {volume}
  {97}},\ \bibinfo {pages} {032311} (\bibinfo {year} {2018})}\BibitemShut
  {NoStop}%
\bibitem [{\citenamefont {Leverrier}(2015)}]{Leverrier2015a}%
  \BibitemOpen
  \bibfield  {author} {\bibinfo {author} {\bibfnamefont {A.}~\bibnamefont
  {Leverrier}},\ }\href {\doibase 10.1103/PhysRevLett.114.070501} {\bibfield
  {journal} {\bibinfo  {journal} {Phys. Rev. Lett.}\ }\textbf {\bibinfo
  {volume} {114}},\ \bibinfo {pages} {070501} (\bibinfo {year}
  {2015})}\BibitemShut {NoStop}%
\bibitem [{\citenamefont {Lapidoth}\ and\ \citenamefont
  {Shamai}(2002)}]{LapidothShamai2002}%
  \BibitemOpen
  \bibfield  {author} {\bibinfo {author} {\bibfnamefont {A.}~\bibnamefont
  {Lapidoth}}\ and\ \bibinfo {author} {\bibfnamefont {S.}~\bibnamefont
  {Shamai}},\ }\href {\doibase 10.1109/18.995552} {\bibfield  {journal}
  {\bibinfo  {journal} {IEEE Transactions on Information Theory}\ }\textbf
  {\bibinfo {volume} {48}},\ \bibinfo {pages} {1118} (\bibinfo {year}
  {2002})}\BibitemShut {NoStop}%
\bibitem [{\citenamefont {Fossier}\ \emph {et~al.}(2009)\citenamefont
  {Fossier}, \citenamefont {Diamanti}, \citenamefont {Debuisschert},
  \citenamefont {Tualle-Brouri},\ and\ \citenamefont {Grangier}}]{Fossier2009}%
  \BibitemOpen
  \bibfield  {author} {\bibinfo {author} {\bibfnamefont {S.}~\bibnamefont
  {Fossier}}, \bibinfo {author} {\bibfnamefont {E.}~\bibnamefont {Diamanti}},
  \bibinfo {author} {\bibfnamefont {T.}~\bibnamefont {Debuisschert}}, \bibinfo
  {author} {\bibfnamefont {R.}~\bibnamefont {Tualle-Brouri}}, \ and\ \bibinfo
  {author} {\bibfnamefont {P.}~\bibnamefont {Grangier}},\ }\href {\doibase
  10.1088/0953-4075/42/11/114014} {\bibfield  {journal} {\bibinfo  {journal}
  {J. Phys. B}\ }\textbf {\bibinfo {volume} {42}},\ \bibinfo {pages} {114014}
  (\bibinfo {year} {2009})}\BibitemShut {NoStop}%
\bibitem [{\citenamefont {Lodewyck}\ \emph {et~al.}(2007)\citenamefont
  {Lodewyck}, \citenamefont {Bloch}, \citenamefont {Garc{\'{i}}a-Patr{\'{o}}n},
  \citenamefont {Fossier}, \citenamefont {Karpov}, \citenamefont {Diamanti},
  \citenamefont {Debuisschert}, \citenamefont {Cerf}, \citenamefont
  {Tualle-Brouri}, \citenamefont {McLaughlin},\ and\ \citenamefont
  {Grangier}}]{Lodewyck2007}%
  \BibitemOpen
  \bibfield  {author} {\bibinfo {author} {\bibfnamefont {J.}~\bibnamefont
  {Lodewyck}}, \bibinfo {author} {\bibfnamefont {M.}~\bibnamefont {Bloch}},
  \bibinfo {author} {\bibfnamefont {R.}~\bibnamefont
  {Garc{\'{i}}a-Patr{\'{o}}n}}, \bibinfo {author} {\bibfnamefont
  {S.}~\bibnamefont {Fossier}}, \bibinfo {author} {\bibfnamefont
  {E.}~\bibnamefont {Karpov}}, \bibinfo {author} {\bibfnamefont
  {E.}~\bibnamefont {Diamanti}}, \bibinfo {author} {\bibfnamefont
  {T.}~\bibnamefont {Debuisschert}}, \bibinfo {author} {\bibfnamefont {N.~J.}\
  \bibnamefont {Cerf}}, \bibinfo {author} {\bibfnamefont {R.}~\bibnamefont
  {Tualle-Brouri}}, \bibinfo {author} {\bibfnamefont {S.~W.}\ \bibnamefont
  {McLaughlin}}, \ and\ \bibinfo {author} {\bibfnamefont {P.}~\bibnamefont
  {Grangier}},\ }\href {\doibase 10.1103/PhysRevA.76.042305} {\bibfield
  {journal} {\bibinfo  {journal} {Phys. Rev. A}\ }\textbf {\bibinfo {volume}
  {76}},\ \bibinfo {pages} {042305} (\bibinfo {year} {2007})}\BibitemShut
  {NoStop}%
\bibitem [{\citenamefont {Soh}\ \emph {et~al.}(2015)\citenamefont {Soh},
  \citenamefont {Brif}, \citenamefont {Coles}, \citenamefont
  {L{\"{u}}tkenhaus}, \citenamefont {Camacho}, \citenamefont {Urayama},\ and\
  \citenamefont {Sarovar}}]{Soh2015}%
  \BibitemOpen
  \bibfield  {author} {\bibinfo {author} {\bibfnamefont {D.~B.~S.}\
  \bibnamefont {Soh}}, \bibinfo {author} {\bibfnamefont {C.}~\bibnamefont
  {Brif}}, \bibinfo {author} {\bibfnamefont {P.~J.}\ \bibnamefont {Coles}},
  \bibinfo {author} {\bibfnamefont {N.}~\bibnamefont {L{\"{u}}tkenhaus}},
  \bibinfo {author} {\bibfnamefont {R.~M.}\ \bibnamefont {Camacho}}, \bibinfo
  {author} {\bibfnamefont {J.}~\bibnamefont {Urayama}}, \ and\ \bibinfo
  {author} {\bibfnamefont {M.}~\bibnamefont {Sarovar}},\ }\href {\doibase
  10.1103/PhysRevX.5.041010} {\bibfield  {journal} {\bibinfo  {journal} {Phys.
  Rev. X}\ }\textbf {\bibinfo {volume} {5}},\ \bibinfo {pages} {041010}
  (\bibinfo {year} {2015})}\BibitemShut {NoStop}%
\bibitem [{\citenamefont {Qi}\ \emph {et~al.}(2015)\citenamefont {Qi},
  \citenamefont {Lougovski}, \citenamefont {Pooser}, \citenamefont {Grice},\
  and\ \citenamefont {Bobrek}}]{Qi2015}%
  \BibitemOpen
  \bibfield  {author} {\bibinfo {author} {\bibfnamefont {B.}~\bibnamefont
  {Qi}}, \bibinfo {author} {\bibfnamefont {P.}~\bibnamefont {Lougovski}},
  \bibinfo {author} {\bibfnamefont {R.}~\bibnamefont {Pooser}}, \bibinfo
  {author} {\bibfnamefont {W.}~\bibnamefont {Grice}}, \ and\ \bibinfo {author}
  {\bibfnamefont {M.}~\bibnamefont {Bobrek}},\ }\href {\doibase
  10.1103/PhysRevX.5.041009} {\bibfield  {journal} {\bibinfo  {journal} {Phys.
  Rev. X}\ }\textbf {\bibinfo {volume} {5}},\ \bibinfo {pages} {041009}
  (\bibinfo {year} {2015})}\BibitemShut {NoStop}%
\bibitem [{\citenamefont {Ali}\ \emph {et~al.}(1998)\citenamefont {Ali},
  \citenamefont {Al-Dhahir},\ and\ \citenamefont {Hershey}}]{DopplerLEO}%
  \BibitemOpen
  \bibfield  {author} {\bibinfo {author} {\bibfnamefont {I.}~\bibnamefont
  {Ali}}, \bibinfo {author} {\bibfnamefont {N.}~\bibnamefont {Al-Dhahir}}, \
  and\ \bibinfo {author} {\bibfnamefont {J.}~\bibnamefont {Hershey}},\ }\href
  {\doibase 10.1109/26.662636} {\bibfield  {journal} {\bibinfo  {journal}
  {Communications, IEEE Transactions on}\ }\textbf {\bibinfo {volume} {46}},\
  \bibinfo {pages} {309} (\bibinfo {year} {1998})}\BibitemShut {NoStop}%
\bibitem [{\citenamefont {Shoji}\ \emph {et~al.}(2012)\citenamefont {Shoji},
  \citenamefont {Fice}, \citenamefont {Takayama},\ and\ \citenamefont
  {Seeds}}]{Shoji2012}%
  \BibitemOpen
  \bibfield  {author} {\bibinfo {author} {\bibfnamefont {Y.}~\bibnamefont
  {Shoji}}, \bibinfo {author} {\bibfnamefont {M.~J.}\ \bibnamefont {Fice}},
  \bibinfo {author} {\bibfnamefont {Y.}~\bibnamefont {Takayama}}, \ and\
  \bibinfo {author} {\bibfnamefont {A.~J.}\ \bibnamefont {Seeds}},\ }\href
  {\doibase 10.1109/jlt.2012.2204037} {\bibfield  {journal} {\bibinfo
  {journal} {Journal of Lightwave Technology}\ }\textbf {\bibinfo {volume}
  {30}},\ \bibinfo {pages} {2696} (\bibinfo {year} {2012})}\BibitemShut
  {NoStop}%
\bibitem [{\citenamefont {Paillier}\ \emph {et~al.}(2019)\citenamefont
  {Paillier}, \citenamefont {Conan}, \citenamefont {Bidan}, \citenamefont
  {Artaud}, \citenamefont {Vedrenne},\ and\ \citenamefont
  {Jaouen}}]{Paillier2019}%
  \BibitemOpen
  \bibfield  {author} {\bibinfo {author} {\bibfnamefont {L.}~\bibnamefont
  {Paillier}}, \bibinfo {author} {\bibfnamefont {J.-M.}\ \bibnamefont {Conan}},
  \bibinfo {author} {\bibfnamefont {R.~L.}\ \bibnamefont {Bidan}}, \bibinfo
  {author} {\bibfnamefont {G.}~\bibnamefont {Artaud}}, \bibinfo {author}
  {\bibfnamefont {N.}~\bibnamefont {Vedrenne}}, \ and\ \bibinfo {author}
  {\bibfnamefont {Y.}~\bibnamefont {Jaouen}},\ }in\ \href {\doibase
  10.1109/icsos45490.2019.8978983} {\emph {\bibinfo {booktitle} {2019 {IEEE}
  International Conference on Space Optical Systems and Applications
  ({ICSOS})}}}\ (\bibinfo  {publisher} {{IEEE}},\ \bibinfo {year}
  {2019})\BibitemShut {NoStop}%
\bibitem [{\citenamefont {Gao}\ \emph {et~al.}(2015)\citenamefont {Gao},
  \citenamefont {Peng}, \citenamefont {Zhang}, \citenamefont {Evariste},
  \citenamefont {Liu}, \citenamefont {Wang}, \citenamefont {Zhong},
  \citenamefont {Lin}, \citenamefont {Wang}, \citenamefont {Chen},\ and\
  \citenamefont {Xu}}]{Gao2015}%
  \BibitemOpen
  \bibfield  {author} {\bibinfo {author} {\bibfnamefont {F.}~\bibnamefont
  {Gao}}, \bibinfo {author} {\bibfnamefont {B.}~\bibnamefont {Peng}}, \bibinfo
  {author} {\bibfnamefont {Y.}~\bibnamefont {Zhang}}, \bibinfo {author}
  {\bibfnamefont {N.~H.}\ \bibnamefont {Evariste}}, \bibinfo {author}
  {\bibfnamefont {J.}~\bibnamefont {Liu}}, \bibinfo {author} {\bibfnamefont
  {X.}~\bibnamefont {Wang}}, \bibinfo {author} {\bibfnamefont {M.}~\bibnamefont
  {Zhong}}, \bibinfo {author} {\bibfnamefont {M.}~\bibnamefont {Lin}}, \bibinfo
  {author} {\bibfnamefont {N.}~\bibnamefont {Wang}}, \bibinfo {author}
  {\bibfnamefont {R.}~\bibnamefont {Chen}}, \ and\ \bibinfo {author}
  {\bibfnamefont {H.}~\bibnamefont {Xu}},\ }\href {\doibase
  10.1016/j.asr.2014.11.032} {\bibfield  {journal} {\bibinfo  {journal}
  {Advances in Space Research}\ }\textbf {\bibinfo {volume} {55}},\ \bibinfo
  {pages} {1394} (\bibinfo {year} {2015})}\BibitemShut {NoStop}%
\bibitem [{\citenamefont {Kuchynka}\ \emph {et~al.}(2020)\citenamefont
  {Kuchynka}, \citenamefont {Serrano}, \citenamefont {Merz},\ and\
  \citenamefont {Siminski}}]{Kuchynka2020}%
  \BibitemOpen
  \bibfield  {author} {\bibinfo {author} {\bibfnamefont {P.}~\bibnamefont
  {Kuchynka}}, \bibinfo {author} {\bibfnamefont {M.~M.}\ \bibnamefont
  {Serrano}}, \bibinfo {author} {\bibfnamefont {K.}~\bibnamefont {Merz}}, \
  and\ \bibinfo {author} {\bibfnamefont {J.}~\bibnamefont {Siminski}},\ }\href
  {\doibase 10.1017/aer.2020.8} {\bibfield  {journal} {\bibinfo  {journal} {The
  Aeronautical Journal}\ }\textbf {\bibinfo {volume} {124}},\ \bibinfo {pages}
  {888} (\bibinfo {year} {2020})}\BibitemShut {NoStop}%
\bibitem [{\citenamefont {Leverrier}\ \emph {et~al.}(2010)\citenamefont
  {Leverrier}, \citenamefont {Grosshans},\ and\ \citenamefont
  {Grangier}}]{Leverrier2010}%
  \BibitemOpen
  \bibfield  {author} {\bibinfo {author} {\bibfnamefont {A.}~\bibnamefont
  {Leverrier}}, \bibinfo {author} {\bibfnamefont {F.}~\bibnamefont
  {Grosshans}}, \ and\ \bibinfo {author} {\bibfnamefont {P.}~\bibnamefont
  {Grangier}},\ }\href {\doibase 10.1103/PhysRevA.81.062343} {\bibfield
  {journal} {\bibinfo  {journal} {Phys. Rev. A}\ }\textbf {\bibinfo {volume}
  {81}},\ \bibinfo {pages} {062343} (\bibinfo {year} {2010})}\BibitemShut
  {NoStop}%
\end{thebibliography}%

\end{document}